\documentclass[lettersize,journal]{IEEEtran}
\usepackage{amsmath,amsfonts}
\usepackage{algorithmic}
\usepackage{algorithm}
\usepackage{array}
\usepackage[caption=false,font=normalsize,labelfont=sf,textfont=sf]{subfig}
\usepackage{textcomp}
\usepackage{stfloats}
\usepackage{url}
\usepackage{verbatim}
\usepackage{graphicx}
\usepackage{cite}
\usepackage{color}
\usepackage{cuted}\stripsep -3pt plus 3pt minus 2pt
\usepackage{lipsum}
\usepackage{esint}

\begin{document}

\title{Block-Level Interference Exploitation Precoding for MU-MISO: An ADMM Approach}
\author{Yiran Wang, Yunsi Wen, Ang Li,~\IEEEmembership{Senior Member,~IEEE}, Xiaoyan Hu,~\IEEEmembership{Member,~IEEE}, and Christos Masouros,~\IEEEmembership{Senior Member,~IEEE}}

\maketitle
\begin{abstract}
We study constructive interference based block-level precoding (CI-BLP) in the downlink of multi-user multiple-input single-output (MU-MISO) systems.
Specifically, our aim is to extend the analysis on CI-BLP to the case where the considered number of symbol slots is smaller than that of the users. To this end, we mathematically prove the feasibility of using the pseudo-inverse to obtain the optimal CI-BLP precoding matrix in a closed form. Similar to the case when the number of users is small, we show that a quadratic programming (QP) optimization on simplex can be constructed.
We also design a low-complexity algorithm based on the alternating direction method of multipliers (ADMM) framework, which can efficiently solve large-scale QP problems. We further analyze the convergence and complexity of the proposed algorithm.
Numerical results validate our analysis and the optimality of the proposed algorithm, and further show that the proposed algorithm offers a flexible performance-complexity tradeoff by limiting the maximum number of iterations, which motivates the use of CI-BLP in practical wireless systems. 
\end{abstract}


\section{Introduction}
\IEEEPARstart{P}{recoding} has been widely studied in multiple-input multiple-output (MIMO) communication system, which is able to support data transmission to multiple users simultaneously \cite{1}.
In the downlink, if the channel state information (CSI) is fully known to the base station, dirty paper coding (DPC) can achieve the best performance by pre-subtracting interference before transmission \cite{2}, but its prohibitive computational costs make it difficult to implement in practical systems.
Therefore, low-complexity closed-form linear precoding schemes, represented by ZF \cite{5} and RZF \cite{6}, are proposed to reduce the computational complexity in signal processing.
At the same time, optimization-based precoding schemes are gaining more and more attention because they allow precoding to better meet various communication constraints and requirements in different scenarios.
One popular example is the downlink signal-to-interference-plus-noise ratio (SINR) balancing approach, which aims to achieve a desired SINR for each user under transmit power constraints \cite{7}.
Another popular form is to minimize the transmitted power under the SINR constraint of each user \cite{8}.
\cite{9} proves that the SINR balancing and the power minimization are dual problem to each other, where an effective iterative algorithm is proposed by exploiting such duality to efficiently solve these two problems.

More recent research has shown that multi-user interference need not be completely eliminated. This is because interference can be utilized by interference exploitation precoding techniques to benefit symbol detection, thus improving the error-rate performance of MIMO communication systems.
In \cite{28}, the concept of 'constructive interference' (CI) is introduced, and CI-based precoding has received increasing research attention.
In \cite{10}, multi-user interference is strictly aligned with the desired data symbol, and a CI-based maximum ratio transmission (MRT) precoding design is carried out to achieve improved performance. This approach was later shown to be sub-optimal and referred to as the ‘strict phase-rotation’ CI metric.
The concept of ‘constructive region’ is introduced in \cite{11} and \cite{12}, which shows that CI does not have to be strictly aligned with the desired data symbol and as long as the interfered signals lie in the constructive region, the effect of interference is construtive.
This observation alleviates the requirement that the interfering signals have to be strictly rotated to the direction of the intended data symbols, leading to further performance improvements. The CI metric introduced in \cite{11} was later named the ‘non-strict phase-rotation’ CI metric and is widely adopted in the relevant literature. 
Meanwhile, a relaxed CI metric based on a ‘relaxed detection region’ was introduced in \cite{12}, which expands the constructive region based on a phase margin that is related to the signal-to-noise ratio (SNR) target. The above CI-based precoding approaches are all designed for PSK modulation, while \cite{29} was the first to extend the exploitation of CI to QAM modulation, where the CI effect can be exploited by the outer constellation points of a QAM constellation by employing the ‘symbol-scaling’ CI metric.
Due to the significant advantages of CI, symbol level precoding based on CI has been applied to intelligent reflecting surface (IRS)-assisted communication \cite{13}, \cite{14}, 1-bit precoding \cite{15}, \cite{16}, \cite{17}, \cite{18}, radar-communication coexistence \cite{19}, \cite{20} and many other wireless communication scenarios.

It has to be mentioned that the utilization of CI in the above schemes is based on symbol level precoding, which requires the base station to solve a different CI-based symbol-level precoding (CI-SLP) optimization problem for each symbol slot.
Symbol-by-symbol optimization brings significant computational burden to the signal processing unit and requires high real-time processing capability.
To alleviate the computational costs, several studies attempt to reduce the complexity of the CI-SLP optimization problem, including derivations of the optimal precoding structure of CI-SLP with efficient iterative algorithms \cite{22}, \cite{26}, sub-optimal solutions \cite{31}, \cite{32}, and deep learning-based methods \cite{33}, \cite{34}, \cite{35}. Specifically, \cite{22} and \cite{26} derive the optimal precoding structure of CI-SLP for PSK and QAM modulation, respectively, and show that the CI-SLP optimization problem can equivalently be transformed into a QP optimization problem and solved using an iterative algorithm with a closed-form solution at each step. Building upon this, the work in \cite{31} derives an exact closed-form but sub-optimal solution for the power minimization CI-SLP problem. 
Despite the above attempts to reduce the computational costs of solving the CI-SLP optimization problem for each symbol slot, these approaches still require solving an optimization problem at the symbol level, i.e., the total number of CI-SLP optimization problems that needs to be solved in a channel coherence interval is not reduced. 
In order to further motivate the realization of CI-based precoding techniques in practical communication wireless systems, \cite{21} proposed CI-based block-level precoding (CI-BLP) for multi-user multiple-input single-output (MU-MISO) communication system for the first time.
Based on Lagrange function and KKT condition, the closed-form expression of CI-BLP optimal precoding matrix is derived when the number of symbol slots in a block is not smaller than the number of users. By further studying the corresponding duality problem, the original optimization problem is transformed into a quadratic programming (QP) optimization on simplex.

However, it is still unclear whether a similar result exists for the case when the number of symbols in the considered block is smaller than the number of users.
This is because the results in \cite{21} cannot be directly applied to the above scenarios, which will be shown mathematically in this paper.
Moreover, despite the fact that CI-BLP and the traditional CI-SLP method have similar problem structure, it is still unclear whether the iterative closed-form algorithm proposed in CI-SLP \cite{22} can offer complexity benefits when used to solve the QP problem for CI-BLP.

We aim to extend the analysis on CI-BLP to the case where the number of symbol slots in the considered block is smaller than that of the users, and propose efficient algorithms to obtain the optimal CI-BLP  precoding matrix. For clarity, we summarize the main contributions of the paper below:

\begin{itemize}

\item[1)] We extend the analysis on CI-BLP to the case where the number of symbol slots in the considered block is smaller than that of the users, where the results in \cite{21} are not directly applicable. Specifically, we mathematically prove the feasibility of using the pseudo-inverse to obtain the optimal CI-BLP precoding matrix in a closed form. Building upon this, a similar QP optimization on simplex can also be constructed, and we thus unify the precoding structure of CI-BLP for all possible communication scenarios.
\item[2)] To validate whether the existing closed-form iterative algorithms designed for CI-SLP can be directly applied to CI-BLP,We study the rank of the quadratic coefficient matrix of the formulated QP problem for CI-BLP mathematically. Since the iterative closed-form algorithm requires the quadratic coefficient matrix to be invertible, we show that the iterative closed-form algorithm is applicable only when the number of symbol slots in a block is smaller than the number of users. However, the convergence rate is not promising for the QP problems of CI-BLP.
\item[3)] Therefore, to design efficient algorithms to obtain the optimal precoding matrix for CI-BLP, we leverage the ADMM framework and we obtain the closed-form solution of each subproblem within the ADMM iteration. In addition, based on the equivalent transformation of the original QP problem, an improved ADMM algorithm is proposed. We analyze the convergence and complexity of this algorithm. Compared with the original ADMM algorithm, the updated variables can be obtained in a simpler way in the improved ADMM algorithm, and better performance can be obtained with fewer iterations.
\item[4)] 
Simulation results show that the proposed precoding structure of CI-BLP can get the same result as the original optimization problem solved by CVX.
Despite taking the simplest initialization values and parameters, the proposed ADMM algorithm can get satisfactory results in dozens of iterations. And it is much faster than the traditional interior point method (IPM) \cite{25}.
More importantly, the proposed ADMM algorithm is more time-efficient than \emph{quadprog}, which motivates the use of the block-level CI beamforming in practice.

\end{itemize}

The remainder of this paper is organized as follows.
Section \uppercase\expandafter{\romannumeral2} introduces the system model and briefly reviews the problem fomulation of CI-BLP.
Section \uppercase\expandafter{\romannumeral3} reviews the CI-BLP problems, and Section \uppercase\expandafter{\romannumeral4} extends the analysis of CI-BLP to the scenario when the number of symbol slots in the considered block is smaller than the number of users.
The discussion on whether the iterative closed-form algorithm in CI-SLP can be used directly in CI-BLP is given in Section \uppercase\expandafter{\romannumeral5}, and the proposed iterative closed-form algorithm based on ADMM is introduced in Section \uppercase\expandafter{\romannumeral6}. Numerical results are provided in Section \uppercase\expandafter{\romannumeral7}, and Section \uppercase\expandafter{\romannumeral8} concludes the paper.

\textit{Notations:} Herein, Lowercase, boldface lowercase and boldface uppercase letters denote scalars, vectors and matrices, respectively. $\mathbb{R}$ and $\mathbb{C}$ denote the set of real numbers and the set of complex numbers, respectively. Superscripts $^{\rm T}$ denotes the transpose. The operator $\|\cdot\|_{2}$ denotes the $2$-norm of a vector. $\mathfrak{R}\left\{\cdot\right\}$ and $\mathfrak{I}\left\{\cdot\right\}$ extract the real and imaginary parts of the argument, respectively. We use $\Pi_{\boldsymbol{\Omega}}$ to represent the projection of the argument onto the set $\boldsymbol{\Omega}$. The relational operators $\preceq$ and $\succeq$ describe element-wise inequality relationships. We define $j$ as $\sqrt{-1}$ and $\mathbf{I}_{N}$ as the $N\times N$ identity matrix. Finally, $\mathbf{1}_{N}$ and $\mathbf{0}_{N}$ denote the $N\times 1$ all-one vector and all-zero vector, respectively.

\section{System Model and Preliminaries}
\subsection{System Model}
We consider the generic multi-user multiple-input single-output (MU-MISO) downlink system, where a base station (BS) equipped with $N_{t}$ antennas serves $K$ single-antenna users simultaneously. 

For the transmission of a block of symbol slots, the data symbol vector in the $n$-th slot is denoted by $\mathbf{s}^{n}=\left[s^{n}_{1},s^{n}_{2},\cdots,s^{n}_{K}\right]^{\rm T}\in \mathbb{C}^{K}$, which is assumed to be drawn from a unit-norm $\cal{M}$-PSK constellation. Accordingly, the received signal for user $k$ in the $n$-th symbol slot can be expressed as
\begin{equation}
\label{equation1}
y^{n}_{k}=\mathbf{h}^{\rm T}_{k}\mathbf{W}\mathbf{s}^{n}+z^{n}_{k},
\end{equation}
where $\mathbf{h}_{k}\in\mathbb{C}^{{N}_{t}}$ is the channel between BS and user $k$, which is constant within the considered block, and $z^{n}_{k}\in\mathbb{C}$ is the additive noise of user ${k}$ in the $n$-th symbol slot. $\mathbf{W}\in \mathbb{C}^{N_{t}\times K}$ is the precoding matrix that applies to all $\mathbf{s}^{n}$ in the considered block. $n \in \left\{n|n\leq N \right\}$, where $N$ is the length of the considered transmission block.

\subsection{Preliminaries on the Symbol-Scaling CI Metric and CI-BLP Optimization Problem}
CI can increase the useful signal power and greatly improve the performance of multi-user transmission. 
To illustrate the symbol-scaling CI meric introduced in \cite{27}, below we depict one quarter of an 8PSK constellation in Fig. 1. 
Without loss of generality, we assume that $\overrightarrow{OA}$ is the nominal constellation point for user $k$ in the $n$-th slot, i.e.,
\begin{equation}
\label{equation2}
\overrightarrow{OA}=s^{n}_{k}.
\end{equation}
$\overrightarrow{OB}$ represents the noiseless received signal with interference, where based on the geometry we obtain
\begin{align}
\label{equation3}
\overrightarrow{OB}&=\overrightarrow{OA}+\overrightarrow{AB}\notag\\
&=\mathbf{h}^{n}_{k}\mathbf{W}\mathbf{s}^{n},
\end{align}
where $\overrightarrow{AB}$ can be regarded as the interference from other user streams.
\begin{figure}[h]
\centering
\includegraphics[width=3in]{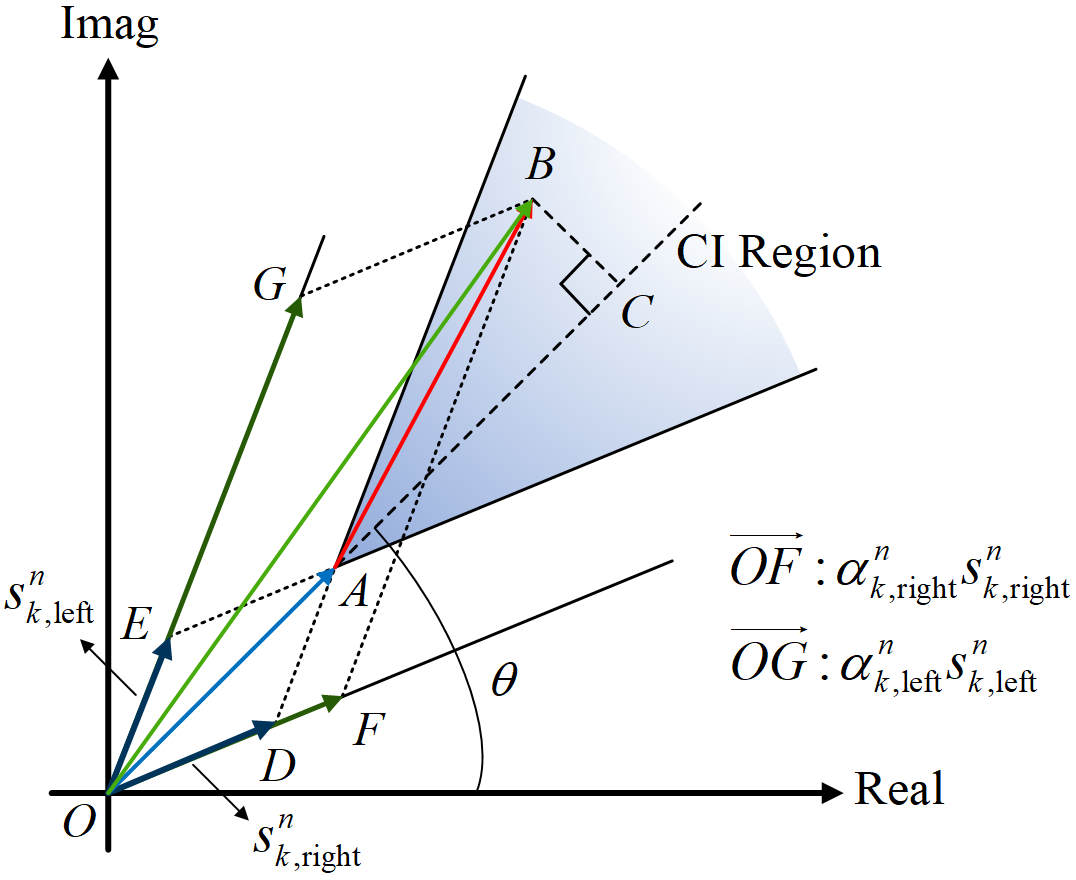}
\caption{Geometric diagram of the symbol-scaling CI metric for 8PSK.}
\label{fig_1}
\end{figure}

Different from the common phase-rotation CI metric which uses phase relations, the symbol-scaling CI metric decomposes the signal along the decision boundaries and imposes scaling constraints on the decomposed components. 
In Fig. 1, $\overrightarrow{OA}$ is decomposed along the two decision boundaries for 8PSKmodulation to obtain $\overrightarrow{OD}$ and $\overrightarrow{OE}$:
\begin{align}
\label{equation4}
\overrightarrow{OA}&=\overrightarrow{OD}+\overrightarrow{OE}\notag\\
&=s^{n}_{k,\text{right}}+s^{n}_{k,\text{left}}.
\end{align}
Following a similar procedure, the received signal $\overrightarrow{OB}$ can also be decomposed along the two decision boundaries into
\begin{align}
\label{equation6}
\overrightarrow{OB}&=\overrightarrow{OF}+\overrightarrow{OG}\notag\\
&=\alpha^{n}_{k,\text{right}}s^{n}_{k,\text{right}}+\alpha^{n}_{k,\text{left}}s^{n}_{k,\text{left}}.
\end{align}
By following the transformation in Section \uppercase\expandafter{\romannumeral4}-A of \cite{16}, which we omit in this paper for brevity, we can construct a coefficient matrix $\mathbf{M}^{n} \in \mathbb{R}^{2K \times 2N_{t}}$ and obtain:
\begin{align}
\label{equation8}
\boldsymbol{\alpha}^{n}_{\rm E}=\mathbf{M}^{n}\mathbf{W}_{\rm E}\mathbf{s}^{n}_{\rm E},
\end{align}
where $\boldsymbol{\alpha}^{n}_{\rm E}\in \mathbb{R}^{2K}$, $\mathbf{W}_{\rm E}\in \mathbb{R}^{2N_{t} \times 2K}$ and $\mathbf{s}_{\rm E}^{n}\in \mathbb{R}^{2K}$ are defined as
\begin{align}
\label{equation9}
\boldsymbol{\alpha}^{n}_{\rm E}=\left[\alpha^{n}_{1,\text{right}},\alpha^{n}_{2,\text{right}},\cdots,\alpha^{n}_{K,\text{right}},\alpha^{n}_{1,\text{left}},\alpha^{n}_{2,\text{left}},\cdots,\alpha^{n}_{K,\text{left}}\right]^{\rm T},
\end{align}
\begin{align}
\label{equation10}
\mathbf{W}_{\rm E}=
\begin{bmatrix} 
\mathfrak{R}\left(\mathbf{W}\right) & -\mathfrak{I}\left(\mathbf{W}\right) \\
\mathfrak{I}\left(\mathbf{W}\right) & \mathfrak{R}\left(\mathbf{W}\right) 
\end{bmatrix},
\end{align}
\begin{align}
\label{equation11}
\mathbf{s}^{n}_{\rm E}=\left[\mathfrak{R}\left(\mathbf{s}^{n}\right)^{\rm T},\mathfrak{I}\left(\mathbf{s}^{n}\right)^{\rm T}\right]^{\rm T}.
\end{align}

Recalling Fig. 1, we can observe that the value of $\alpha^{n}_{k,\text{right}}$ or $\alpha^{n}_{k,\text{left}}$ represents the effect of inter-user interference, and a larger value of $\alpha^{n}_{k,\text{right}}$ or $\alpha^{n}_{k,\text{left}}$ means that the symbol is pushed further away from one of its decision boundary. Accordingly, we can then construct the CI-BLP optimization problem
that maximizes the minimum value of the entry in $\boldsymbol{\alpha}^{n}_{\rm E}$ for all the considered symbol slots within the block, given by
\begin{align}
\label{equation12}
\mathcal{P}_{0}:\ \max_{\mathbf{W}_{\rm E}}\ &\min_{k,n}\ \alpha^{n}_{k}\notag\\
\text{s.t.}\ \ &\boldsymbol{\alpha}^{n}_{\rm E}=\mathbf{M}^{n}\mathbf{W}_{\rm E}\mathbf{s}^{n}_{\rm E},\ \forall n\leq N\notag\\
&\sum_{n=1}^{N}\|\mathbf{W}_{\rm E}\mathbf{s}^{n}_{\rm E}\|^{2}_{2}\leq Np_{0}
\end{align}
where $\alpha^{n}_{k}$ represents the $k$-th entry in $\boldsymbol{\alpha}^{n}_{\rm E}$, and $p_{0}$ represents the transmit power budget per symbol slot.

\section{QP optimization for the case of $N\geq K$}
$\mathcal{P}_{0}$ is a joint optimization over all symbol slots within the block, and it is a convex problem that can be directly solved via optimization tools such as CVX. In this section, we introduce the process of transforming $\mathcal{P}_{0}$ into a QP problem for $N\geq K$ in \cite{21}.

To facilitate subsequent derivations, we introduce $\hat{\mathbf{W}}\in \mathbb{R}^{N_{t} \times 2K}$:
\begin{align}
\label{equation13}
\hat{\mathbf{W}}=\left[\mathfrak{R}\left(\mathbf{W}\right)\ -\mathfrak{I}\left(\mathbf{W}\right)\right],
\end{align}
based on which $\mathbf{W}_{\rm E}$ can be expressed as
\begin{align}
\label{equation14}
\mathbf{W}_{\rm E}=\mathbf{P}\hat{\mathbf{W}}+\mathbf{Q}\hat{\mathbf{W}}\mathbf{T},
\end{align}
where $\mathbf{P}\in \mathbb{R}^{2N_{t}\times N_{t}}$, $\mathbf{Q}\in \mathbb{R}^{2N_{t}\times N_{t}}$ and $\mathbf{T}\in \mathbb{R}^{2K\times K}$ are defined as
\begin{align}
\label{equation15}
\mathbf{P}=
\begin{bmatrix}
\mathbf{I}_{N_{t}\times N_{t}}\\
\mathbf{O}_{N_{t}\times N_{t}}
\end{bmatrix}
,\ 
\mathbf{Q}=
\begin{bmatrix}
\mathbf{O}_{N_{t}\times N_{t}}\\
\mathbf{I}_{N_{t}\times N_{t}}
\end{bmatrix}
,\ 
\mathbf{T}=
\begin{bmatrix}
\mathbf{O}_{K\times K} & \mathbf{I}_{K\times K}\\
-\mathbf{I}_{K\times K} & \mathbf{O}_{K\times K}
\end{bmatrix}.
\end{align}
The expression for $\boldsymbol{\alpha}^{n}_{\rm E}$ can be further transformed into:
\begin{align}
\label{equation16}
\boldsymbol{\alpha}^{n}_{\rm E}&=\mathbf{M}^{n}\mathbf{W}_{\rm E}\mathbf{s}^{n}_{\rm E}\notag \\
&=\mathbf{M}^{n}\left(\mathbf{P}\hat{\mathbf{W}}+\mathbf{Q}\hat{\mathbf{W}}\mathbf{T}\right)\mathbf{s}^{n}_{\rm E}\notag \\
&=\mathbf{M}^{n}\mathbf{P}\hat{\mathbf{W}}\mathbf{s}^{n}_{\rm E}+\mathbf{M}^{n}\mathbf{Q}\hat{\mathbf{W}}\mathbf{T}\mathbf{s}^{n}_{\rm E}\notag \\
&=\mathbf{A}^{n}\hat{\mathbf{W}}\mathbf{s}^{n}_{\rm E}+\mathbf{B}^{n}\hat{\mathbf{W}}\mathbf{c}^{n}_{\rm E},
\end{align}
where $\mathbf{A}^{n} \in \mathbb{R}^{2K\times N_{t}}$, $\mathbf{B}^{n}\in \mathbb{R}^{2K\times N_{t}}$, $\mathbf{c}^{n}_{\rm E}\in \mathbb{R}^{2K}$ are defined as
\begin{align}
\label{equation17}
\mathbf{A}^{n}=\mathbf{M}^{n}\mathbf{P}
, \ 
\mathbf{B}^{n}=\mathbf{M}^{n}\mathbf{Q}
, \ 
\mathbf{c}^{n}_{\rm E}=\mathbf{T}\mathbf{s}^{n}_{\rm E}.
\end{align}
Accordingly, the $k$-th entry of $\boldsymbol{\alpha}^{n}$ can be expressed as
\begin{align}
\label{equation18}
\alpha^{n}_{k}=\left(\mathbf{a}^{n}_{k}\right)^{\rm T}\hat{\mathbf{W}}\mathbf{s}^{n}_{\rm E}+\left(\mathbf{b}^{n}_{k}\right)^{\rm T}\hat{\mathbf{W}}\mathbf{s}^{n}_{\rm E},
\end{align}
where $\left(\mathbf{a}^{n}_{k}\right)^{\rm T}$ and $\left(\mathbf{b}^{n}_{k}\right)^{\rm T}$ represent the $k$-th row of $\mathbf{A}^{n}$ and $\mathbf{B}^{n}$, respectively. With the above expression, $\mathcal{P}_{0}$ can be expressed as a standard convex optimization problem:
\begin{align}
\label{equation19}
\mathcal{P}_{1}:\ \min_{\hat{\mathbf{W}}, t}\ &-t\notag\\
\text{s.t.}\ \ &t-\left(\mathbf{a}^{n}_{k}\right)^{\rm T}\hat{\mathbf{W}}\mathbf{s}^{n}_{\rm E}-\left(\mathbf{b}^{n}_{k}\right)^{\rm T}\hat{\mathbf{W}}\mathbf{s}^{n}_{\rm E}\leq 0,\ \forall k\leq 2K,\ n\leq N\notag \\
&\sum_{n=1}^{N}\|\left(\mathbf{P}\hat{\mathbf{W}}+\mathbf{Q}\hat{\mathbf{W}}\mathbf{T}\right)\mathbf{s}^{n}_{\rm E}\|^{2}_{2}\leq Np_{0}
\end{align}

By analyzing the Lagrangian and the KKT conditions of $\mathcal{P}_{1}$, we can obtain
\begin{align}
\label{equation20}
2\mu\hat{\mathbf{W}}\mathbf{D}=\sum_{n=1}^{N}\left[\left(\mathbf{A}^{n}\right)^{\rm T}\boldsymbol{\delta}^{n}\left(\mathbf{s}^{n}_{\rm E}\right)^{\rm T}+\left(\mathbf{B}^{n}\right)^{\rm T}\boldsymbol{\delta}^{n}\left(\mathbf{c}^{n}_{\rm E}\right)^{\rm T}\right],
\end{align}
where $\boldsymbol{\delta}^{n}=\left[\delta^{n}_{1},\delta^{n}_{2},\cdots,\delta^{n}_{2K}\right]^{\rm T}$ and $\mu$ are the non-negative dual variables associated with two inequality constraints of $\mathcal{P}_{1}$ respectively, and $\mathbf{D} \in \mathbb{R}^{2K \times 2K}$ is given by
\begin{align}
\label{equation21}
\mathbf{D}=\left[\sum_{n=1}^{N}\mathbf{s}^{n}_{\rm E}\left(\mathbf{s}^{n}_{\rm E}\right)^{\rm T}+\sum_{n=1}^{N}\mathbf{c}^{n}_{\rm E}\left(\mathbf{c}^{n}_{\rm E}\right)^{\rm T}\right].
\end{align}
When $N\geq K$, $\mathbf{D}$ is thus full-rank and invertible. Accordingly, we can obtain an expression for the optimal precoding matrix $\hat{\mathbf{W}}$ in a closed form as
\begin{align}
\label{equation22}
\hat{\mathbf{W}}=\frac{1}{2\mu}\sum_{n=1}^{N}\left[\left(\mathbf{A}^{n}\right)^{\rm T}\boldsymbol{\delta}^{n}\left(\mathbf{s}^{n}_{\rm E}\right)^{\rm T}+\left(\mathbf{B}^{n}\right)^{\rm T}\boldsymbol{\delta}^{n}\left(\mathbf{c}^{n}_{\rm E}\right)^{\rm T}\right]\mathbf{D}^{-1}.
\end{align}
Since problem $\mathcal{P}_{1}$ satisfies Slater condition, we can find the optimal solution of $\mathcal{P}_{1}$ by solving its dual problem. By substituting the closed-form structure of $\hat{\mathbf{W}}$ into the objective function of the dual problem, and by defining $\boldsymbol{\delta}_{\rm E} \in \mathbb{R}^{2NK\times 1}$ and $\mathbf{U}_{m,n} \in \mathbb{R}^{2K \times 2K}$ as
\begin{align}
\label{equation23}
\boldsymbol{\delta}_{\rm E}=\left[\left(\boldsymbol{\delta}^{1}\right)^{\rm T},\left(\boldsymbol{\delta}^{2}\right)^{\rm T},\cdots,\left(\boldsymbol{\delta}^{N}\right)^{\rm T}\right]^{\rm T},
\end{align}
\begin{align}
\label{equation24}
\mathbf{U}_{m,n}=&p_{m,n}\mathbf{A}^{m}\left(\mathbf{A}^{n}\right)^{\rm T}+f_{m,n}\mathbf{A}^{m}\left(\mathbf{B}^{n}\right)^{\rm T}\notag \\
&+g_{m,n}\mathbf{B}^{m}\left(\mathbf{A}^{n}\right)^{\rm T}+q_{m,n}\mathbf{B}^{m}\left(\mathbf{B}^{n}\right)^{\rm T},
\end{align}
where $m \in \left\{1,\cdots,N\right\}$ and $n \in \left\{1,\cdots,N\right\}$. $p_{m,n}$, $f_{m,n}$, $g_{m,n}$ and $q_{m,n}$ are defined as
\begin{align}
\label{equation25}
p_{m,n}=\left(\mathbf{s}^{n}_{\rm E}\right)^{\rm T}\mathbf{D}^{-1}\mathbf{s}^{m}_{\rm E}, \ 
f_{m,n}=\left(\mathbf{c}^{n}_{\rm E}\right)^{\rm T}\mathbf{D}^{-1}\mathbf{s}^{m}_{\rm E}, \notag \\
g_{m,n}=\left(\mathbf{s}^{n}_{\rm E}\right)^{\rm T}\mathbf{D}^{-1}\mathbf{c}^{m}_{\rm E}, \ 
q_{m,n}=\left(\mathbf{c}^{n}_{\rm E}\right)^{\rm T}\mathbf{D}^{-1}\mathbf{c}^{m}_{\rm E},
\end{align}
the final dual problem of CI-BLP can be formulated as 
\begin{align}
\label{equation26}
\mathcal{P}_{2}:\ \min_{\boldsymbol{\delta}_{\rm E}}\ &\boldsymbol{\delta}^{\rm T}_{\rm E}\mathbf{U}\boldsymbol{\delta}_{\rm E}\notag\\
\text{s.t.}\ \ &\mathbf{1}^{\rm T}\boldsymbol{\delta}_{\rm E}-1=0 \notag \\
&\delta^{m}_{\rm E} \geq0,\ \forall m \in\left\{1,2,\cdots,2NK\right\}
\end{align}
where $\delta^{m}_{\rm E}$ is the $ m$-th entry of vector $\boldsymbol{\delta}_{\rm E}$, and $\mathbf{U} \in \mathbb{R}^{2NK \times 2NK}$ is a block matrix constructed as
\begin{align}
\label{equation27}
\mathbf{U}=
\begin{bmatrix}
\mathbf{U}_{1,1} & \mathbf{U}_{1,2} & \cdots & \mathbf{U}_{1,N} \\
\mathbf{U}_{2,1} & \mathbf{U}_{2,2} & \cdots & \mathbf{U}_{2,N} \\
\vdots & \vdots & \ddots & \vdots \\
\mathbf{U}_{N,1} & \mathbf{U}_{N,2} & \cdots & \mathbf{U}_{N,N} \\
\end{bmatrix}.
\end{align}
$\mathcal{P}_{2}$ is a QP optimization problem over a simplex, which can be more efficiently solved than $\mathcal{P}_{1}$. After solving $\mathcal{P}_{2}$ and obtaining $\hat{\mathbf{W}}$ via the closed-form structure of $\hat{\mathbf{W}}$, the original complex precoding matrix $\mathbf{W}$ can be obtained.

\section{QP optimization for the case of $N < K$}
We further extend this study to the case where the number of users that the BS simultaneously serves is larger than the number of symbol slots within the considered transmission block, i.e., $N<K$. 

When $N<K$, the direct inverse included in \eqref{equation22} becomes infeasible, as the matrix $\mathbf{D}$ is rank-deficient. In this case, the more general pseudo inverse instead of the direct matrix inverse is employed. Based on \eqref{equation20}, we can now express $\hat{\mathbf{W}}$ in the case of $N<K$ as
\begin{align}
\label{equation28}
\hat{\mathbf{W}}=\frac{1}{2\mu}\sum_{n=1}^{N}\left[\left(\mathbf{A}^{n}\right)^{\rm T}\boldsymbol{\delta}^{n}\left(\mathbf{s}^{n}_{\rm E}\right)^{\rm T}+\left(\mathbf{B}^{n}\right)^{\rm T}\boldsymbol{\delta}^{n}\left(\mathbf{c}^{n}_{\rm E}\right)^{\rm T}\right]\mathbf{D}^{+}.
\end{align}
Then, one can easily follow a similar approach to that in section \uppercase\expandafter{\romannumeral3} to obtain a QP optimization and the corresponding
solution. Although we note that a pseudo-inverse does not always guarantee the equality of the original constraint, our derivations in this section and the corresponding numerical results show that, in our problem the pseudo-inverse guarantees the equality of the original constraint and the closed-form structure of $\hat{\mathbf{W}}$ given by \eqref{equation28} is feasible.

In order to make the expression more concise, we can define a matrix $\mathbf{C} \in \mathbb{R}^{N_{t} \times 2K}$ as
\begin{align}
\label{equation29}
\mathbf{C}=\frac{1}{2\mu}\sum_{n=1}^{N}\left[\left(\mathbf{A}^{n}\right)^{\rm T}\boldsymbol{\delta}^{n}\left(\mathbf{s}^{n}_{\rm E}\right)^{\rm T}+\left(\mathbf{B}^{n}\right)^{\rm T}\boldsymbol{\delta}^{n}\left(\mathbf{c}^{n}_{\rm E}\right)^{\rm T}\right].
\end{align}
Accordingly, \eqref{equation28} and \eqref{equation20} can be respectively rewritten as
\begin{align}
\label{equation30}
\hat{\mathbf{W}}=\mathbf{C}\mathbf{D}^{+},
\end{align}
\begin{align}
\label{equation31}
\hat{\mathbf{W}}\mathbf{D}=\mathbf{C}.
\end{align}
Obviously, the closed-form structure of $\hat{\mathbf{W}}$ given by \eqref{equation28} is feasible as long as the following relation can be proved:
\begin{align}
\label{equation32}
\hat{\mathbf{W}}=\mathbf{C}\mathbf{D}^{+} \Rightarrow \hat{\mathbf{W}}\mathbf{D}=\mathbf{C}.
\end{align}
In what follows, we analyze the above relation to further obtain the equivalent condition.

According to \eqref{equation21}, $\mathbf{D}$ is a real symmetric matrix. Thus, by eigen decomposition, we can obtain
\begin{align}
\label{equation33}
\mathbf{D}&=\mathbf{V}_{\rm D}\boldsymbol{\Sigma}_{\rm D}\mathbf{V}_{\rm D}^{\rm T}\notag \\
&=\mathbf{V}_{\rm D}
\begin{bmatrix}
\sigma_{1} & & & & & \\
& \ddots & & & & \\
& & \sigma_{r_d} & & & \\
& & & 0 & & \\
& & & & \ddots & \\
& & & & & 0\\
\end{bmatrix}
\mathbf{V}_{\rm D}^{\rm T},
\end{align}
\begin{align}
\label{equation34}
\mathbf{D}^{+}&=\mathbf{V}_{\rm D}\boldsymbol{\Sigma}_{\rm D}^{+}\mathbf{V}_{\rm D}^{\rm T}\notag \\
&=\mathbf{V}_{\rm D}
\begin{bmatrix}
\frac{1}{\sigma_{1}} & & & & & \\
& \ddots & & & & \\
& & \frac{1}{\sigma_{r_d}} & & & \\
& & & 0 & & \\
& & & & \ddots & \\
& & & & & 0\\
\end{bmatrix}
\mathbf{V}_{\rm D}^{\rm T},
\end{align}
where $\boldsymbol{\Sigma}_{\rm D}$ is a diagonal matrix whose diagonal elements are eigenvalues of $\mathbf{D}$, and $r_{d}$ is the rank of $\mathbf{D}$, i.e., number of non-zero eigenvalues of the matrix $\mathbf{D}$. $\mathbf{V}_{\rm D} \in \mathbb{R}^{2K \times 2K}$ is an orthogonal matrix composed of eigenvectors of $\mathbf{D}$.
By substituting \eqref{equation34} into $\mathbf{D}^{+}$ in \eqref{equation30} and substituting \eqref{equation33} into $\mathbf{D}$ in \eqref{equation31}, we can obtain
\begin{align}
\label{equation35}
&\hat{\mathbf{W}}=\mathbf{C}\mathbf{D}^{+} \notag \\
\Leftrightarrow &\hat{\mathbf{W}}=\mathbf{C}\mathbf{V}_{\rm D}\boldsymbol{\Sigma}_{\rm D}^{+}\mathbf{V}_{\rm D}^{\rm T} \notag \\
\Leftrightarrow &\hat{\mathbf{W}}\mathbf{V}_{\rm D}=\mathbf{C}\mathbf{V}_{\rm D}\boldsymbol{\Sigma}_{\rm D}^{+},
\end{align}
\begin{align}
\label{equation36}
&\hat{\mathbf{W}}\mathbf{D}=\mathbf{C} \notag \\
\Leftrightarrow &\hat{\mathbf{W}}\mathbf{V}_{\rm D}\boldsymbol{\Sigma}_{\rm D}\mathbf{V}_{\rm D}^{\rm T}=\mathbf{C} \notag \\
\Leftrightarrow &\hat{\mathbf{W}}\mathbf{V}_{\rm D}\boldsymbol{\Sigma}_{\rm D}=\mathbf{C}\mathbf{V}_{\rm D}.
\end{align}
Accordingly, the feasibility condition equivalent to \eqref{equation32} can be written as
\begin{align}
\label{equation37}
\mathbf{F}=\mathbf{E}\boldsymbol{\Sigma}_{\rm D}^{+} \Rightarrow \mathbf{F}\boldsymbol{\Sigma}_{\rm D}=\mathbf{E},
\end{align}
where $\mathbf{E}$ and $\mathbf{F}$ are defined as
\begin{align}
\label{equation38}
\mathbf{E}=\mathbf{C}\mathbf{V}_{\rm D},\ \mathbf{F}=\hat{\mathbf{W}}\mathbf{V}_{\rm D}.
\end{align}
In order to prove the above equivalent feasibility conditions, the relation of each row needs to be proved. The relation in the $i$-th row can be expanded by element as follows
\begin{align}
\label{equation39}
F_{i,1}=E_{i,1}\cdot\frac{1}{\lambda_{1}} &\Rightarrow F_{i,1}\cdot\lambda{1}=E_{i,1}\notag\\
&\vdots\notag\\
F_{i,r_{d}}=E_{i,r_{d}}\cdot\frac{1}{\lambda_{r_{d}}} &\Rightarrow F_{i,r_{d}}\cdot\lambda{r_{d}}=E_{i,r_{d}}\notag\\
F_{i,r_{d}+1}=E_{i,r_{d}+1}\cdot0 &\Rightarrow F_{i,r_{d}+1}\cdot0=E_{i,r_{d}+1}\notag\\
&\vdots\notag\\
F_{i,2K}=E_{i,2K}\cdot0 &\Rightarrow F_{i,2K}\cdot0=E_{i,2K}\notag\\
\end{align}
where $i \in \left\{1,\cdots,N_{t}\right\}$ and $m \in \left\{1,\cdots,2K\right\}$. $E_{i,m}$ and $F_{i,m}$ represent the elements of the $i$-th row and $m$-th column of matrix $\mathbf{E}$ and matrix $\mathbf{F}$ respectively.
The first $r_{d}$ expressions in \eqref{equation40} are naturally true, and the equivalent condition for the remaining expressions is
\begin{align}
\label{equation40}
E_{i,r_{d}+1}=\cdots=E_{i,2K}=0.
\end{align}
By decomposing $\mathbf{E}=\left[\mathbf{e}^{1},\mathbf{e}^{2},\cdots,\mathbf{e}^{2K}\right]$, the equivalent feasibility condition of \eqref{equation32} can be rewritten as
\begin{align}
\label{equation41}
\mathbf{e}^{r_D+1}=\cdots=\mathbf{e}^{2K}=\mathbf{0}.
\end{align}

To proceed, we figure out the properties of a matrix that is constructed in a particular way.

\newtheorem{proposition}{\bf Proposition}
\begin{proposition}\label{prop1}
Suppose $\left[\mathbf{s}_{\rm G}^{1},\mathbf{s}_{\rm G}^{2},\cdots,\mathbf{s}_{\rm G}^{N_{g}}\right] \in \mathbb{R}^{M_{g} \times N_{g}}$. $\text{rank}\left(\left[\mathbf{s}_{\rm G}^{1},\mathbf{s}_{\rm G}^{2},\cdots,\mathbf{s}_{\rm G}^{N_{g}}\right]\right)=r_{g}$, and $\mathbf{s}_{\rm G}^{1},\mathbf{s}_{\rm G}^{2},\cdots,\mathbf{s}_{\rm G}^{r_{g}}$ are linearly independent. Then we can construct a matrix 
\begin{align}
\label{equation42}
\mathbf{G}&=\sum_{n=1}^{N_{g}}\mathbf{s}^{n}_{\rm G}\left(\mathbf{s}^{n}_{\rm G}\right)^{\rm T} \notag \\
&=\mathbf{V}_{\rm G}\boldsymbol{\Sigma}_{\rm G}\mathbf{V}_{\rm G}^{\rm T} \notag \\
&=\mathbf{V}_{\rm G}
\begin{bmatrix}
\tilde{\sigma}_{1} & & & & & \\
& \ddots & & & & \\
& & \tilde{\sigma}_{r_{g}} & & & \\
& & & 0 & & \\
& & & & \ddots & \\
& & & & & 0\\
\end{bmatrix}
\mathbf{V}_{\rm G}^{\rm T}
\end{align}
where $\mathbf{V}_{\rm G}=\left[\mathbf{v}_{\rm G}^{1},\mathbf{v}_{\rm G}^{2},\cdots,\mathbf{v}_{\rm G}^{M_{g}}\right]$. 
For any $n\in\left\{1,2,\cdots,N_{g}\right\},\ m\in\left\{r_{g}+1,r_{g}+2,\cdots,M_{g}\right\}$, the following condition must be satisfied:
\begin{align}
\label{equation43}
\left(\mathbf{s}_{\rm G}^{n}\right)^{\rm T}\cdot\mathbf{v}_{\rm G}^{m}=0.
\end{align}
\end{proposition}

{\bf{Proof:}}See Appendix A.


Proposition 1 shows that for matrix $\mathbf{D}$ given in \eqref{equation21},
The following conditions must be satisfied:
\begin{align}
\label{equation52}
&\left(\mathbf{s}_{\rm E}^{n}\right)^{\rm T}\cdot\mathbf{v}_{\rm D}^{m}=0,\ 
\left(\mathbf{c}_{\rm E}^{n}\right)^{\rm T}\cdot\mathbf{v}_{\rm D}^{m}=0,
\end{align}
where $n \in \left\{1,\cdots,N\right\},\ m \in \left\{r_{d}+1,\cdots,2K\right\}$, and $\mathbf{v}_{\rm D}^{m}$ is the $m$-th column of matrix $\mathbf{V}_{\rm D}$. 

Accordingly, for any $m \in \left\{r_d+1,\cdots,2K\right\}$,
\begin{align}
\label{equation53}
\mathbf{e}^{m}&=\left(\frac{1}{2\mu}\sum_{n=1}^{N}\left[\left(\mathbf{A}^{n}\right)^{\rm T}\boldsymbol{\delta}^{n}\left(\mathbf{s}^{n}_{\rm E}\right)^{\rm T}+\left(\mathbf{B}^{n}\right)^{\rm T}\boldsymbol{\delta}^{n}\left(\mathbf{c}^{n}_{\rm E}\right)^{\rm T}\right]\right)\cdot\mathbf{v}_{\rm D}^{m}\notag \\
&=\frac{1}{2\mu}\sum_{n=1}^{N}\left[\left(\mathbf{A}^{n}\right)^{\rm T}\boldsymbol{\delta}^{n}\left(\mathbf{s}^{n}_{\rm E}\right)^{\rm T}\cdot\mathbf{v}_{\rm D}^{m}+\left(\mathbf{B}^{n}\right)^{\rm T}\boldsymbol{\delta}^{n}\left(\mathbf{c}^{n}_{\rm E}\right)^{\rm T}\cdot\mathbf{v}_{\rm D}^{m}\right]\notag \\
&=\frac{1}{2\mu}\sum_{n=1}^{N}\left[\left(\mathbf{A}^{n}\right)^{\rm T}\boldsymbol{\delta}^{n}\cdot 0+\left(\mathbf{B}^{n}\right)^{\rm T}\boldsymbol{\delta}^{n}\cdot 0\right]\notag \\
&=\mathbf{0}
\end{align}
which means that the equivalent feasibility condition in \eqref{equation41} is satisfied and the closed-form structure of $\hat{\mathbf{W}}$ given by \eqref{equation28} is feasible.

We note that when $N\geq K$, $\mathbf{D}$ is full-rank and the closed-form structure of $\hat{\mathbf{W}}$ given by \eqref{equation28} is also feasible. Therefore, we can obtain a uniform closed-form structure of $\hat{\mathbf{W}}$ and we can design optimal block-level precoding by solving $\mathcal{P}_{2}$ for systems of any size.

\section{Application of iterative closed-form scheme}
$\mathcal{P}_{2}$ is a QP optimization problem over a simplex, which can be more efficiently solved by the iterative closed-form scheme proposed in \cite{22}. Compared with the interior point method, the closed iteration scheme is less complex, and initialization and parameter selection are not required. However, the quadratic coefficient matrix in the QP problem is required to be invertible in this iterative closed-form scheme. Therefore, before applying this iterative closed-form scheme to $\mathcal{P}_{2}$, the rank of $\mathbf{U}$ needs to be discussed.

\subsection{Rank of $\mathbf{U}$}
Before analyzing the rank of $\mathbf{U}$, we state that the variables introduced in the proof of each subsequent proposition are only used to complete the proof and have nothing to do with other variables of the same name in this paper.

According to \eqref{equation24} and \eqref{equation26}, we can  further get a new expression of $\mathbf{U}$ as shown in \eqref{equation54}.
We can further define

In addition, in order to derive the rank of $\mathbf{U}$, we also define matrix $\hat{\mathbf{U}}$, matrix $\hat{\mathbf{U}}_{1}$, and matrix $\hat{\mathbf{U}}_{2}$ as \eqref{equation55}, \eqref{equation56} and \eqref{equation57} respectively. The new expression of $\mathbf{U}$ and the definitions of $\hat{\mathbf{U}}$, $\hat{\mathbf{U}}_{1}$ and $\hat{\mathbf{U}}_{2}$ are in Appendix B.

\begin{proposition}\label{prop2}
$\text{rank}\left(\mathbf{D}\right)=\text{min}\left\{2K,2N\right\}$.
\end{proposition}

{\bf{Proof:}}See Appendix C.


\begin{proposition}\label{prop3}
$\text{rank}\left(\hat{\mathbf{U}}_{1}\right)=\text{rank}\left(\mathbf{D}\right)=\text{min}\left\{2K,2N\right\}$.
\end{proposition}

{\bf{Proof:}}See Appendix D.


\begin{proposition}\label{prop4}
$\text{rank}\left(\hat{\mathbf{U}}_{2}\right)=K$.
\end{proposition}

{\bf{Proof:}}See Appendix E.


\begin{proposition}\label{prop5}
$\text{rank}\left(\hat{\mathbf{U}}\right)=\text{min}\left\{2NK,2K^{2}\right\}$.
\end{proposition}

{\bf{Proof:}}See Appendix F.


\begin{proposition}\label{prop6}
$\text{rank}\left(\mathbf{U}\right)=\text{min}\left\{2NK,2K^{2}\right\}$.
\end{proposition}

{\bf{Proof:}}See Appendix G.


\subsection{Iterative Closed-Form Scheme}
The rank of $\mathbf{U} \in \mathbb{R}^{2NK \times 2NK}$ is $\text{min}\left\{2NK,2K^{2}\right\}$, which means that $\mathbf{U}$ is full rank when $N<K$.
Since the quadratic coefficient matrix in the QP problem is required to be invertible in iterative closed-form scheme \cite{22}, we can only use the iterative closed form scheme when $N<K$.

We simulate systems of different scales using iterative closed algorithm, and obtaine their corresponding cumulative distribution functions (CDF) of maximum iterations respectively, as shown in Figure 2.
The selected system sizes are:
\begin{itemize}
\item[1)]$N_{t}=K=6, N=4$;
\item[2)]$N_{t}=K=8, N=6$;
\item[3)]$N_{t}=K=10, N=8$;
\item[4)]$N_{t}=K=12, N=10$.
\end{itemize}

\begin{figure}[h]
\centering
\includegraphics[width=3.5in]{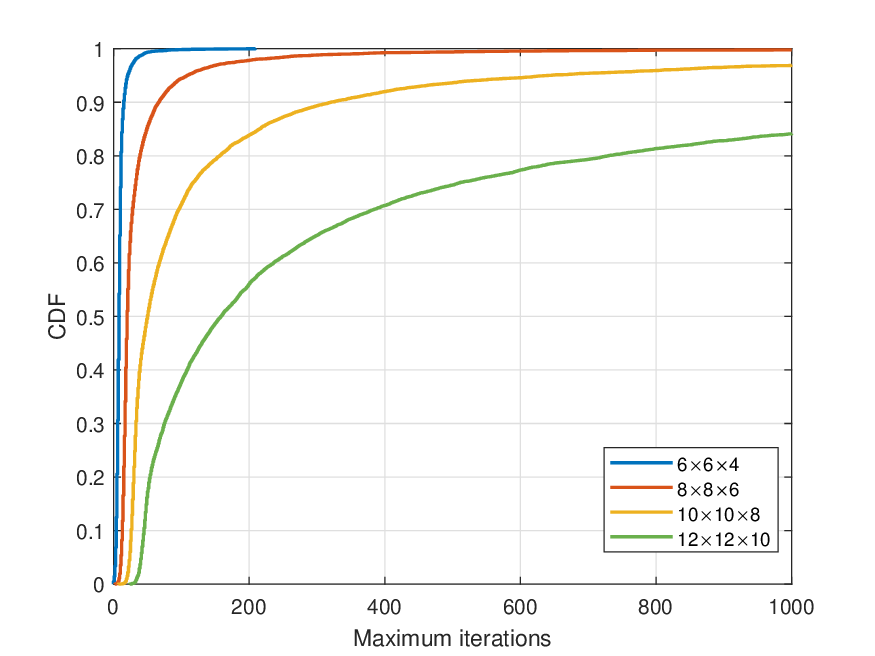}
\caption{CDF of maximum iterations.}
\label{fig_2}
\end{figure}

We found that as the size of the problem increased, the number of iterations required was quite large. Therefore, iterative closed form algorithm are not suitable for large-scale QP problems. In addition, the iterative closed form algorithm can only be applied to the scenario when $N<K$. The iterative closed-form solution is not ideal for solving $\mathcal{P}_{2}$. Therefore, we hope to use ADMM to design a low complexity scheme to solve $\mathcal{P}_{2}$.

\section{ADMM Solution to Solve $\mathcal{P}_{2}$}
In this section, we present two ADMM schemes to solve $\mathcal{P}_{2}$. For the second ADMM scheme, we give the convergence analysis and the method to accelerate the convergence speed.

\subsection{The ADMM algorithm based on $\mathcal{P}_{2}$}
To apply the ADMM framework, we introduce a new variable $\boldsymbol{\omega}$ and define a set $\Omega=\left\{\boldsymbol{\omega}|\omega_{i}\geq 0,\ \forall i\in\left\{1,2,\dots,2NK\right\}\right\}$. $\mathcal{P}_{2}$ can be written as
\begin{align}
\label{equation1-1}
\mathcal{P}_{3}:\ \min_{\boldsymbol{\delta}_{\rm E},\boldsymbol{\omega}}\ &f\left(\boldsymbol{\delta}_{\rm E}\right)\notag\\
\text{s.t.}\ \ &\mathbf{1}^{\rm T}\boldsymbol{\delta}_{\rm E}-1=0\notag \\
\ \ &\boldsymbol{\delta}_{\rm E}-\boldsymbol{\omega}=\mathbf{0}\notag\\
\ \ &\boldsymbol{\omega} \in \Omega
\end{align}
By defining an indicator function for $\boldsymbol{\omega}$

\begin{align}
\label{equation1-2}
I_{\Omega}\left(\boldsymbol{\omega}\right) = \begin{cases}
0,&{\text{if}}\ \boldsymbol{\omega}\in \Omega, \\ 
{\infty,}&{\text{otherwise}} .
\end{cases}
\end{align}
the problem can be rewritten in an equivalent consensus form as follows \cite{23}
\begin{align}
\label{equation1-3}
\mathcal{P}_{4}:\ \min_{\boldsymbol{\delta}_{\rm E},\boldsymbol{\omega}}\ &f\left(\boldsymbol{\delta}_{\rm E}\right)+I_{\Omega}\left(\boldsymbol{\omega}\right)\notag\\
\text{s.t.}\ \ &\boldsymbol{\delta}_{\rm E}-\boldsymbol{\omega}=\mathbf{0}
\end{align}
where $f\left(\boldsymbol{\delta}_{\rm E}\right)$ is defined as
\begin{align}
\label{equation1-4}
f\left(\boldsymbol{\delta}_{\rm E}\right)=\boldsymbol{\delta}_{\rm E}^{\rm T}\mathbf{U}\boldsymbol{\delta}_{\rm E},\ \ \ {\rm \mathbf{dom}}\ f=\left\{\boldsymbol{\delta}_{\rm E}|\mathbf{1}^{\rm T}\boldsymbol{\delta}_{\rm E}-1=0\right\} 
\end{align}
For the optimization problem $\mathcal{P}_{4}$, its augmented Lagrangian
is expressed as
\begin{align}
\label{equation1-5}
L_{\rho}\left(\boldsymbol{\delta}_{\rm E},\boldsymbol{\omega},\boldsymbol{\lambda}\right)&=\boldsymbol{\delta}_{\rm E}^{\rm T}\mathbf{U}\boldsymbol{\delta}_{\rm E}+I_{\Omega}\left(\boldsymbol{\omega}\right)-\left<\boldsymbol{\lambda},\boldsymbol{\delta}_{\rm E}-\boldsymbol{\omega}\right>+\frac{\rho}{2}\left\|\boldsymbol{\delta}_{\rm E}-\boldsymbol{\omega}\right\|_{2}^{2}\notag\\
&=\boldsymbol{\delta}_{\rm E}^{\rm T}\mathbf{U}\boldsymbol{\delta}_{\rm E}+I_{\Omega}\left(\boldsymbol{\omega}\right)+\frac{\rho}{2}\left\|\boldsymbol{\delta}_{\rm E}-\boldsymbol{\omega}-\frac{\boldsymbol{\lambda}}{\rho}\right\|_{2}^{2}-\frac{\left\|\boldsymbol{\lambda}\right\|_{2}^{2}}{2\rho}
\end{align}
where $\boldsymbol{\lambda}$ is the dual vector and $\rho>0$ is the penalty parameter.
The ADMM framework for the update of $\boldsymbol{\delta}_{\rm E}$, $\boldsymbol{\omega}$, and $\boldsymbol{\lambda}$ can be written as
\begin{align}
\boldsymbol{\delta}_{\rm E}^{k+1}&=\underset{\boldsymbol{\delta}_{\rm E}\in\left\{\boldsymbol{\delta}_{\rm E}|\mathbf{1}^{\rm T}\boldsymbol{\delta}_{\rm E}-1=0\right\}}{\text{arg min}}L_{\rho}\left(\boldsymbol{\delta}_{\rm E},\boldsymbol{\omega}^{k},\boldsymbol{\lambda}^{k}\right),\label{equation1-6}\\
\boldsymbol{\omega}^{k+1}&=\ \ \ \ \underset{\boldsymbol{\omega}}{\text{arg min}}\ \ \ \ L_{\rho}\left(\boldsymbol{\delta}_{\rm E}^{k+1},\boldsymbol{\omega},\boldsymbol{\lambda}^{k}\right),\label{equation1-7}\\
\boldsymbol{\lambda}^{k+1}&=\boldsymbol{\lambda}^{k}-{\rho}\left(\boldsymbol{\delta}_{\rm E}^{k+1}-\boldsymbol{\omega}^{k+1}\right).\label{equation1-8}
\end{align}

In the $\boldsymbol{\delta}_{\rm E}$-update, the optimization problem for $\boldsymbol{\delta}_{\rm E}$ can be written as
\begin{align}
\label{equation1-9}
\min_{\boldsymbol{\delta}_{\rm E}}\ \ &\boldsymbol{\delta}_{\rm E}^{\rm T}\mathbf{U}\boldsymbol{\delta}_{\rm E}+\frac{\rho}{2}\left\|\boldsymbol{\delta}_{\rm E}-\boldsymbol{\omega}^{k}-\frac{\boldsymbol{\lambda}^{k}}{\rho}\right\|_{2}^{2}\notag\\
\text{s.t.}\ \ &\mathbf{1}^{\rm T}\boldsymbol{\delta}_{\rm E}-1=0
\end{align}
For the case of a convex quadratic function restricted to an affine set. 
We can obtain the KKT (Karush-Kuhn-Tucker) condition of this optimization problem:
\begin{align}
\mathbf{1}^{\rm T}\boldsymbol{\delta}_{\rm E}-1=0\label{equation1-10}\\
2\mathbf{U}\boldsymbol{\delta}_{\rm E}+\rho\left(\boldsymbol{\delta}_{\rm E}-\boldsymbol{\omega}^{k}-\frac{\boldsymbol{\lambda}^{k}}{\rho}\right)+\nu\cdot\mathbf{1}=\mathbf{0}\label{equation1-11}
\end{align}
where $\nu$ is the dual variable associated with the equality constraint. Then we can obtain 
\begin{align}
\label{equation1-12}
\begin{bmatrix} 
2\mathbf{U}+\rho & \mathbf{1} \\
\mathbf{1}^{\rm T} & {0}
\end{bmatrix}
\begin{bmatrix} 
\boldsymbol{\delta}_{\rm E} \\
\nu 
\end{bmatrix}
=
\begin{bmatrix} 
\rho\boldsymbol{\omega}^{k}+\boldsymbol{\lambda}^{k} \\
1
\end{bmatrix}
\end{align}
With the above formulation, we can obtain $\boldsymbol{\delta}_{\rm E}^{k+1}$ by an inverse operation
\begin{align}
\label{equation1-13}
\begin{bmatrix} 
\boldsymbol{\delta}_{\rm E}^{k+1} \\
\nu 
\end{bmatrix}
=
\begin{bmatrix} 
2\mathbf{U}+\rho & \mathbf{1} \\
\mathbf{1}^{\rm T} & {0}
\end{bmatrix}
^{-1}
\begin{bmatrix} 
\rho\boldsymbol{\omega}^{k}+\boldsymbol{\lambda}^{k} \\
1
\end{bmatrix}
\end{align}

In the $\boldsymbol{\omega}$-update, the optimization problem for $\boldsymbol{\omega}$ can be written as
\begin{align}
\label{equation1-14}
\min_{\boldsymbol{\omega}}\ \ &I_{\Omega}\left(\boldsymbol{\omega}\right)+\frac{\rho}{2}\left\|\boldsymbol{\delta}_{\rm E}^{k+1}-\boldsymbol{\omega}-\frac{\boldsymbol{\lambda}^{k}}{\rho}\right\|_{2}^{2}
\end{align}
Then $\boldsymbol{\omega}$ needs to satisfy
\begin{align}
\label{equation1-15}
\rho\left(\boldsymbol{\delta}_{\rm E}^{k+1}-\boldsymbol{\omega}-\frac{\boldsymbol{\lambda}^{k}}{\rho}\right) \in \partial I_{\Omega}\left(\boldsymbol{\omega}\right)
\end{align}
and we can obtain
\begin{align}
\label{equation1-16}
\boldsymbol{\omega}^{k+1}={\text max}\left\{\mathbf{0},\boldsymbol{\delta}_{\rm E}^{k+1}-\frac{\boldsymbol{\lambda}^{k}}{\rho}\right\}
\end{align}
The corresponding algorithm is summarized in Algorithm 1.
\begin{algorithm}[H]
\caption{The first ADMM scheme}\label{alg:algorithm1}
\begin{algorithmic}[1] 
\STATE  \textbf{Input}: $\mathbf{s}$, $\mathbf{H}$
\STATE  \textbf{Initial}: $\boldsymbol{\delta}_{\rm E}^{1}=\boldsymbol{\omega}^{1}=\boldsymbol{\lambda}^{1}=\mathbf{0}$, $\mathbf{U}$, $\rho$, maximum number of iterations $K_{max}$.
\STATE  \textbf{for} $k=1,\cdots K_{max}$ \textbf{do} 
\STATE  \hspace{0.5cm} Compute $\boldsymbol{\delta}_{\rm E}^{k+1}$ by \eqref{equation1-13}
\STATE  \hspace{0.5cm} Compute $\boldsymbol{\omega}^{k+1}$ by \eqref{equation1-16}
\STATE  \hspace{0.5cm} Update $\boldsymbol{\lambda}^{k+1}$ by \eqref{equation1-8}
\STATE  \textbf{end for}
\STATE  \textbf{Output}: $\boldsymbol{\delta}_{\rm E}=\boldsymbol{\delta}_{\rm E}^{K_{max}+1}$
\end{algorithmic}
\label{algorithm1}
\end{algorithm}

\subsection{The ADMM algorithm based on $\mathcal{P}_{5}$}
Since the updating process of $\boldsymbol{\delta}_{\rm E}$ in $\mathcal{P}_{4}$ is a constrained optimization problem, the convergence of ADMM algorithm requires a certain number of iterations. We hope to accelerate the convergence speed of ADMM algorithm to reduce the complexity of the algorithm.
We find that $\mathcal{P}_{2}$ is equivalent to the following optimization problem
\begin{align}
\label{equation2-17}
\mathcal{P}_{5}:\ \min_{\boldsymbol{\delta}_{\rm E}}\ &\boldsymbol{\delta}_{\rm E}^{\rm T}\mathbf{U}\boldsymbol{\delta}_{\rm E}\notag\\
\text{s.t.}\ \ &\mathbf{1}^{\rm T}\boldsymbol{\delta}_{\rm E}-1\geq0\notag\\
\ \ &\delta_{\rm E}^{m}\geq0,\ \forall m \in\left\{1,2,\cdots,2NK\right\}
\end{align}
Assume that $\boldsymbol{\delta}^{*}_{\rm E}$ is an optimal solution to $\mathcal{P}_{5}$, such that
\begin{align}
\mathbf{1}^{\rm T}\boldsymbol{\delta}^{*}_{\rm E}-1>0,\label{equation2-18}\\
\left(\delta_{\rm E}^{*}\right)^{m}\geq0,\ \forall m \in\left\{1,2,\cdots,2NK\right\}.\label{equation2-19}
\end{align}
Then there must be $\boldsymbol{\delta}^{**}_{\rm E}=\kappa\boldsymbol{\delta}^{*}_{\rm E}$, such that
\begin{align}
\mathbf{1}^{\rm T}\boldsymbol{\delta}^{**}_{\rm E}-1=0,\label{equation2-20}\\
\left(\delta_{\rm E}^{**}\right)^{m}\geq0,\ \forall m \in\left\{1,2,\cdots,2NK\right\}.\label{equation2-21}
\end{align}
where $\kappa=\frac{1}{\mathbf{1}^{\rm T}\boldsymbol{\delta}^{*}_{\rm E}}\in\left(0,1\right)$. And we have
\begin{align}
\label{equation2-22}
\left(\boldsymbol{\delta}^{**}_{\rm E}\right)^{\rm T}\mathbf{U}\boldsymbol{\delta}^{**}_{\rm E}=\kappa^{2}\left(\boldsymbol{\delta}^{*}_{\rm E}\right)^{\rm T}\mathbf{U}\boldsymbol{\delta}^{*}_{\rm E}<\left(\boldsymbol{\delta}^{*}_{\rm E}\right)^{\rm T}\mathbf{U}\boldsymbol{\delta}^{*}_{\rm E}
\end{align}
which contradicts the assumption that $\boldsymbol{\delta}^{*}_{\rm E}$ is an optimal solution to $\mathcal{P}_{5}$. Therefore, the optimal solution to $\mathcal{P}_{5}$ must satisfy $\mathbf{1}^{\rm T}\boldsymbol{\delta}_{\rm E}-1=0$, which means that the optimal solution to $\mathcal{P}_{5}$ is the optimal solution to $\mathcal{P}_{2}$, that is, $\mathcal{P}_{5}$ is equivalent to $\mathcal{P}_{2}$.

Furthermore, to apply the ADMM framework, we introduce a new variable $\hat{\boldsymbol{\omega}}$ and define a set $\hat{\Omega}=\left\{\hat{\boldsymbol{\omega}}|\hat{\omega}_{i}\geq 0,\ \forall i\in\left\{1,2,\dots,2NK+1\right\}\right\}$. $\mathcal{P}_{5}$ is equivalent to
\begin{align}
\label{equation2-23}
\mathcal{P}_{6}:\ \min_{\boldsymbol{\delta}_{\rm E}}\ &\boldsymbol{\delta}_{\rm E}^{\rm T}\mathbf{U}\boldsymbol{\delta}_{\rm E}\notag\\
\text{s.t.}\ \ &
\begin{bmatrix}
\mathbf{1}^{\rm T}\\
\mathbf{I}_{2NK \times 2NK}
\end{bmatrix}
\boldsymbol{\delta}_{\rm E}
=
\begin{bmatrix}
1\\
\mathbf{0}
\end{bmatrix}
+\hat{\boldsymbol{\omega}}\notag\\
\ \ &\hat{\boldsymbol{\omega}}\in \hat{\Omega}
\end{align}
By defining an indicator function for $\hat{\boldsymbol{\omega}}$
\begin{align}
\label{equation2-24}
I_{\hat{\Omega}}\left(\hat{\boldsymbol{\omega}}\right) = \begin{cases}
0,&{\text{if}}\ \hat{\boldsymbol{\omega}}\in \hat{\Omega}, \\ 
{\infty,}&{\text{otherwise}} .
\end{cases}
\end{align}
and
\begin{align}
\label{equation2-25}
\boldsymbol{\Gamma}=
\begin{bmatrix}
\mathbf{1}^{\rm T}\\
\mathbf{I}_{2NK \times 2NK}
\end{bmatrix}\in\mathbb{C}^{\left(2NK+1\right)\times\left(2NK\right)},\ 
\mathbf{c}=
\begin{bmatrix}
1\\
\mathbf{0}
\end{bmatrix}\in\mathbb{C}^{\left(2NK+1\right)\times1}
\end{align}
$\mathcal{P}_{6}$ can be written as
\begin{align}
\label{equation2-26}
\mathcal{P}_{7}:\ \min_{\boldsymbol{\delta}_{\rm E}}\ &\boldsymbol{\delta}_{\rm E}^{\rm T}\mathbf{U}\boldsymbol{\delta}_{\rm E}+I_{\hat{\Omega}}\left(\hat{\boldsymbol{\omega}}\right)\notag\\
\text{s.t.}\ \ &\boldsymbol{\Gamma}\boldsymbol{\delta}_{\rm E}=\mathbf{c}
+\hat{\boldsymbol{\omega}}
\end{align}
The corresponding augmented Lagrangian function for $\mathcal{P}_{7}$ is expressed as
\begin{align}
\label{equation2-27}
\hat{L}_{\rho}\left(\boldsymbol{\delta}_{\rm E},\hat{\boldsymbol{\omega}},\hat{\boldsymbol{\lambda}}\right)&=\boldsymbol{\delta}_{\rm E}^{\rm T}\mathbf{U}\boldsymbol{\delta}_{\rm E}+I_{\hat{\Omega}}\left(\hat{\boldsymbol{\omega}}\right)+\hat{\boldsymbol{\lambda}}^{\rm T}\left(-\boldsymbol{\Gamma}\boldsymbol{\delta}_{\rm E}+\mathbf{c}+\hat{\boldsymbol{\omega}}\right)\notag\\
&\ \ \ +\frac{\rho}{2}\left\|-\boldsymbol{\Gamma}\boldsymbol{\delta}_{\rm E}+\mathbf{c}+\hat{\boldsymbol{\omega}}\right\|_{2}^{2}
\notag\\
&=\boldsymbol{\delta}_{\rm E}^{\rm T}\mathbf{U}\boldsymbol{\delta}_{\rm E}+I_{\hat{\Omega}}\left(\hat{\boldsymbol{\omega}}\right)\notag\\
&\ \ \ +\frac{\rho}{2}\left\|-\boldsymbol{\Gamma}\boldsymbol{\delta}_{\rm E}+\mathbf{c}+\hat{\boldsymbol{\omega}}+\frac{\hat{\boldsymbol{\lambda}}}{\rho}\right\|_{2}^{2}-\frac{\left\|\hat{\boldsymbol{\lambda}}\right\|_{2}^{2}}{2\rho}
\end{align}
where $\boldsymbol{\lambda}\in \mathbb{C}^{\left(2NK+1\right)\times1}$ is the dual vector and $\rho>0$ is the penalty parameter.
The ADMM framework for the update of $\boldsymbol{\delta}_{\rm E}$, $\hat{\boldsymbol{\omega}}$, and $\hat{\boldsymbol{\lambda}}$ can be written as
\begin{align}
\boldsymbol{\delta}_{\rm E}^{k+1}&=\underset{\boldsymbol{\delta}_{\rm E}}{\text{arg min}}\ \hat{L}_{\rho}\left(\boldsymbol{\delta}_{\rm E},\hat{\boldsymbol{\omega}}^{k},\hat{\boldsymbol{\lambda}}^{k}\right),\label{equation2-28}\\
\hat{\boldsymbol{\omega}}^{k+1}&=\ \underset{\boldsymbol{\omega}}{\text{arg min}}\ \hat{L}_{\rho}\left(\boldsymbol{\delta}_{\rm E}^{k+1},\hat{\boldsymbol{\omega}},\hat{\boldsymbol{\lambda}}^{k}\right),\label{equation2-29}\\
\hat{\boldsymbol{\lambda}}^{k+1}&=\hat{\boldsymbol{\lambda}}^{k}+{\rho}\left(-\boldsymbol{\Gamma}\boldsymbol{\delta}_{\rm E}^{k+1}+\mathbf{c}+\hat{\boldsymbol{\omega}}^{k+1}\right).\label{equation2-30}
\end{align}

In the $\boldsymbol{\delta}_{\rm E}$-update, the optimization problem for $\boldsymbol{\delta}_{\rm E}$ can be written as
\begin{align}
\label{equation2-31}
\min_{\boldsymbol{\delta}_{\rm E}}\ \ \boldsymbol{\delta}_{\rm E}^{\rm T}\mathbf{U}\boldsymbol{\delta}_{\rm E}+\frac{\rho}{2}\left\|-\boldsymbol{\Gamma}\boldsymbol{\delta}_{\rm E}+\mathbf{c}+\hat{\boldsymbol{\omega}}^{k}+\frac{\hat{\boldsymbol{\lambda}}^{k}}{\rho}\right\|_{2}^{2}
\end{align}
For this unconstrained convex optimization problem, the optimal point $\boldsymbol{\delta}_{\rm E}^{t+1}$ should satisfy the condition that the gradient is zero, which leads to 
\begin{align}
\label{equation2-32}
\left(2\mathbf{U}+\rho\boldsymbol{\Gamma}^{\rm T}\boldsymbol{\Gamma}\right)\boldsymbol{\delta}_{\rm E}=\rho\boldsymbol{\Gamma}^{\rm T}\left(\mathbf{c}+\hat{\mathbf{w}}^{k}+\frac{\hat{\boldsymbol{\lambda}}^{k}}{\rho}\right)
\end{align}
Then we can get a closed-form solution for $\boldsymbol{\delta}_{\rm E}^{k+1}$
\begin{align}
\label{equation2-33}
\boldsymbol{\delta}_{\rm E}^{k+1}=\left(2\mathbf{U}+\rho\boldsymbol{\Gamma}^{\rm T}\boldsymbol{\Gamma}\right)^{-1}\rho\boldsymbol{\Gamma}^{\rm T}\left(\mathbf{c}+\hat{\mathbf{w}}^{k}+\frac{\hat{\boldsymbol{\lambda}}^{k}}{\rho}\right)
\end{align}

In the $\boldsymbol{\omega}$-update, the optimization problem for $\boldsymbol{\omega}$ can be written as
\begin{align}
\label{equation2-34}
\min_{\boldsymbol{\omega}}\ \ &I_{\Omega}\left(\boldsymbol{\omega}\right)+\frac{\rho}{2}\left\|-\boldsymbol{\Gamma}\boldsymbol{\delta}_{\rm E}^{k+1}+\mathbf{c}+\hat{\boldsymbol{\omega}}+\frac{\hat{\boldsymbol{\lambda}}^{k}}{\rho}\right\|_{2}^{2}
\end{align}
Then $\boldsymbol{\omega}$ needs to satisfy
\begin{align}
\label{equation2-35}
-\rho\left(-\boldsymbol{\Gamma}\boldsymbol{\delta}_{\rm E}^{k+1}+\mathbf{c}+\hat{\boldsymbol{\omega}}+\frac{\hat{\boldsymbol{\lambda}}^{k}}{\rho}\right) \in \partial I_{\Omega}\left(\boldsymbol{\omega}\right)
\end{align}
and we can obtain
\begin{align}
\label{equation2-36}
\hat{\boldsymbol{\omega}}^{k+1}={\text max}\left\{\mathbf{0},\boldsymbol{\Gamma}\boldsymbol{\delta}_{\rm E}^{k+1}-\mathbf{c}-\frac{\hat{\boldsymbol{\lambda}}^{k}}{\rho}\right\}
\end{align}
The corresponding algorithm is summarized in Algorithm 2.

\begin{algorithm}[H]
\caption{The second ADMM scheme}\label{alg:algorithm2}
\begin{algorithmic}[1] 
\STATE  \textbf{Input}: $\mathbf{s}$, $\mathbf{H}$
\STATE  \textbf{Initial}: $\boldsymbol{\delta}_{\rm E}^{1}=\hat{\boldsymbol{\omega}}^{1}=\hat{\boldsymbol{\lambda}}^{1}=\mathbf{0}$, $\mathbf{U}$, $\mathbf{c}$, $\boldsymbol{\Gamma}$, $\rho$, maximum number of iterations $K_{max}$.
\STATE  \textbf{for} $k=1,\cdots K_{max}$ \textbf{do} 
\STATE  \hspace{0.5cm} Compute $\boldsymbol{\delta}_{\rm E}^{k+1}$ by \eqref{equation2-33}
\STATE  \hspace{0.5cm} Compute $\hat{\boldsymbol{\omega}}^{k+1}$ by \eqref{equation2-36}
\STATE  \hspace{0.5cm} Update $\hat{\boldsymbol{\lambda}}^{k+1}$ by \eqref{equation2-30}
\STATE  \textbf{end for}
\STATE  \textbf{Output}: $\boldsymbol{\delta}_{\rm E}=\boldsymbol{\delta}_{\rm E}^{K_{max}+1}$
\end{algorithmic}
\label{algorithm2}
\end{algorithm}

\subsection{Convergence Analysis for the Second ADMM Scheme}
In order to prove the convergence of our proposed ADMM scheme and to further design a method to accelerate the convergence rate, we analyze the convergence of the second ADMM scheme as follows.

First, the minimizer $\hat{\boldsymbol{\omega}}^{k+1}$ satisfies
\begin{align}
\label{equation2-37}
\hat{L}_{\rho}\left(\boldsymbol{\delta}_{\rm E}^{k},\hat{\boldsymbol{\omega}}^{k+1},\hat{\boldsymbol{\lambda}}^{k}\right) \leq \hat{L}_{\rho}\left(\boldsymbol{\delta}_{\rm E}^{k},\hat{\boldsymbol{\omega}}^{k},\hat{\boldsymbol{\lambda}}^{k}\right)
\end{align}

It is easy to see that $\hat{L}_{\rho}\left(\boldsymbol{\delta}_{\rm E},\hat{\boldsymbol{\omega}}^{k+1},\hat{\boldsymbol{\lambda}}^{k}\right)$ is strongly convex with respect to $\boldsymbol{\delta}_{\rm E}$.
Because of the property of strongly convex function:
\begin{align}
\label{equation2-38}
&\left(\partial_{\boldsymbol{\delta}_{\rm E,1}} \hat{L}_{\rho}\left(\boldsymbol{\delta}_{\rm E,1},\hat{\boldsymbol{\omega}}^{k+1},\hat{\boldsymbol{\lambda}}^{k}\right)- \partial_{\boldsymbol{\delta}_{\rm E,2}} \hat{L}_{\rho}\left(\boldsymbol{\delta}_{\rm E,2},\hat{\boldsymbol{\omega}}^{k+1},\hat{\boldsymbol{\lambda}}^{k}\right)\right)^{\rm T} \notag\\
&\cdot \left(\boldsymbol{\delta}_{\rm E,1}-\hat{\boldsymbol{\delta}}_{\rm E,2}\right)
\geq m\left\|\boldsymbol{\delta}_{\rm E,1}-\hat{\boldsymbol{\delta}}_{\rm E,2}\right\|^{2}_{2}
\end{align}
where $m$ is the strongly convex parameter of $\hat{L}_{\rho}\left(\boldsymbol{\delta}_{\rm E},\hat{\boldsymbol{\omega}}^{k+1},\hat{\boldsymbol{\lambda}}^{k}\right)$. 

We can calculate that
\begin{align}
\label{equation2-39}
&\partial_{\boldsymbol{\delta}_{\rm E,1}} \hat{L}_{\rho}\left(\boldsymbol{\delta}_{\rm E,1},\hat{\boldsymbol{\omega}}^{k+1},\hat{\boldsymbol{\lambda}}^{k}\right)- \partial_{\boldsymbol{\delta}_{\rm E,2}} \hat{L}_{\rho}\left(\boldsymbol{\delta}_{\rm E,2},\hat{\boldsymbol{\omega}}^{k+1},\hat{\boldsymbol{\lambda}}^{k}\right)\notag\\
&=\left(2\mathbf{U}+\rho\boldsymbol{\Gamma}^{\rm T}\boldsymbol{\Gamma}\right)\left(\boldsymbol{\delta}_{\rm E,1}-\boldsymbol{\delta}_{\rm E,2}\right)
\end{align}
Since
\begin{align}
\label{equation2-40}
\rho\boldsymbol{\Gamma}^{\rm T}\boldsymbol{\Gamma}&=\rho
\begin{bmatrix}
\mathbf{1} & 
\mathbf{I}_{2NK \times 2NK}
\end{bmatrix}
\begin{bmatrix}
\mathbf{1}^{\rm T}\\
\mathbf{I}_{2NK \times 2NK}
\end{bmatrix}\notag\\
&=\rho\mathbf{1}\cdot\mathbf{1}^{\rm T}+\rho\mathbf{I}
\end{align}
where $\rho\mathbf{1}\cdot\mathbf{1}^{\rm T}$ is a semi-positive definite matrix and $\mathbf{U}$ is also a semi-positive definite matrix, we can obtain
\begin{align}
\label{equation2-41}
&\left(\partial_{\boldsymbol{\delta}_{\rm E,1}} \hat{L}_{\rho}\left(\boldsymbol{\delta}_{\rm E,1},\hat{\boldsymbol{\omega}}^{k+1},\hat{\boldsymbol{\lambda}}^{k}\right)- \partial_{\boldsymbol{\delta}_{\rm E,2}} \hat{L}_{\rho}\left(\boldsymbol{\delta}_{\rm E,2},\hat{\boldsymbol{\omega}}^{k+1},\hat{\boldsymbol{\lambda}}^{k}\right)\right)^{\rm T}\notag\\
&\cdot \left(\boldsymbol{\delta}_{\rm E,1}-\boldsymbol{\delta}_{\rm E,2}\right) \notag\\
&= \left(\boldsymbol{\delta}_{\rm E,1}-\boldsymbol{\delta}_{\rm E,2}\right)^{\rm T}\left(2\mathbf{U}+\rho\boldsymbol{\Gamma}^{\rm T}\boldsymbol{\Gamma}\right)\left(\boldsymbol{\delta}_{\rm E,1}-\boldsymbol{\delta}_{\rm E,2}\right)  \notag\\
&= \left(\boldsymbol{\delta}_{\rm E,1}-\boldsymbol{\delta}_{\rm E,2}\right)^{\rm T}\left(2\mathbf{U}+\rho\mathbf{1}\cdot\mathbf{1}^{\rm T}+\rho\mathbf{I}\right)\left(\boldsymbol{\delta}_{\rm E,1}-\boldsymbol{\delta}_{\rm E,2}\right)  \notag\\
&\geq \rho \left\|\boldsymbol{\delta}_{\rm E,1}-\boldsymbol{\delta}_{\rm E,2}\right\|^{2}_{2}
\end{align}
Thus, $m=\rho$ and $\hat{L}_{\rho}\left(\boldsymbol{\delta}_{\rm E},\hat{\boldsymbol{\omega}}^{k+1},\hat{\boldsymbol{\lambda}}^{k}\right)$ is $\rho$-strongly convex with respect to $\boldsymbol{\delta}_{\rm E}$.

By definition of strongly convex function:
\begin{align}
\label{equation2-42}
&\hat{L}_{\rho}\left(\boldsymbol{\delta}_{\rm E,1},\hat{\boldsymbol{\omega}}^{k+1},\hat{\boldsymbol{\lambda}}^{k}\right)\geq \hat{L}_{\rho}\left(\boldsymbol{\delta}_{\rm E,2},\hat{\boldsymbol{\omega}}^{k+1},\hat{\boldsymbol{\lambda}}^{k}\right)\notag\\
&+\left(\partial_{\boldsymbol{\delta}_{\rm E,2}} \hat{L}_{\rho}\left(\boldsymbol{\delta}_{\rm E,2},\hat{\boldsymbol{\omega}}^{k+1},\hat{\boldsymbol{\lambda}}^{k}\right)\right)^{\rm T}\left(\boldsymbol{\delta}_{\rm E,1}-\boldsymbol{\delta}_{\rm E,2}\right)\notag\\
&+\frac{\rho}{2}\left\|\boldsymbol{\delta}_{\rm E,1}-\boldsymbol{\delta}_{\rm E,2}\right\|^{2}_{2}
\end{align}
The minimizer $\boldsymbol{\delta}_{\rm E}^{k+1}$ satisfies
\begin{align}
\label{equation2-43}
&\partial_{\boldsymbol{\delta}_{\rm E}^{k+1}} \hat{L}_{\rho}\left(\boldsymbol{\delta}_{\rm E}^{k+1},\hat{\boldsymbol{\omega}}^{k+1},\hat{\boldsymbol{\lambda}}^{k}\right)=\mathbf{0}
\end{align}
Thus, for any $\boldsymbol{\delta}_{\rm E}^{k}$, the minimizer $\boldsymbol{\delta}_{\rm E}^{k+1}$ satisfies
\begin{align}
\label{equation2-44}
&\hat{L}_{\rho}\left(\boldsymbol{\delta}_{\rm E}^{k+1},\hat{\boldsymbol{\omega}}^{k+1},\hat{\boldsymbol{\lambda}}^{k}\right)\notag\\
&\leq \hat{L}_{\rho}\left(\boldsymbol{\delta}_{\rm E}^{k},\hat{\boldsymbol{\omega}}^{k+1},\hat{\boldsymbol{\lambda}}^{k}\right)-\frac{\rho}{2}\left\|\boldsymbol{\delta}_{\rm E}^{k+1}-\boldsymbol{\delta}_{\rm E}^{k}\right\|^{2}_{2}
\end{align}

Moreover, from the definition of $\hat{L}_{\rho}$ and with the use of \eqref{equation2-30},
we have
\begin{align}
\label{equation2-45}
&\hat{L}_{\rho}\left(\boldsymbol{\delta}_{\rm E}^{k+1},\hat{\boldsymbol{\omega}}^{k+1},\hat{\boldsymbol{\lambda}}^{k+1}\right)
-\hat{L}_{\rho}\left(\boldsymbol{\delta}_{\rm E}^{k+1},\hat{\boldsymbol{\omega}}^{k+1},\hat{\boldsymbol{\lambda}}^{k}\right)\notag\\
&=\frac{1}{\rho}\left\|\hat{\boldsymbol{\lambda}}^{k+1}-\hat{\boldsymbol{\lambda}}^{k}\right\|^{2}_{2}
\end{align}
Then, summing \eqref{equation2-37}, \eqref{equation2-44} and \eqref{equation2-45} yields
\begin{align}
\label{equation2-46}
&\hat{L}_{\rho}\left(\boldsymbol{\delta}_{\rm E}^{k+1},\hat{\boldsymbol{\omega}}^{k+1},\hat{\boldsymbol{\lambda}}^{k+1}\right)
-\hat{L}_{\rho}\left(\boldsymbol{\delta}_{\rm E}^{k},\hat{\boldsymbol{\omega}}^{k},\hat{\boldsymbol{\lambda}}^{k}\right)\notag\\
&\leq \frac{1}{\rho}\left\|\hat{\boldsymbol{\lambda}}^{k+1}-\hat{\boldsymbol{\lambda}}^{k}\right\|^{2}_{2}-\frac{\rho}{2}\left\|\boldsymbol{\delta}_{\rm E}^{k+1}-\boldsymbol{\delta}_{\rm E}^{k}\right\|^{2}_{2}
\end{align}
From \eqref{equation2-32}, the minimizer $\boldsymbol{\delta}_{\rm E}^{k+1}$ satisfies
\begin{align}
\label{equation2-47}
\left(2\mathbf{U}+\rho\boldsymbol{\Gamma}^{\rm T}\boldsymbol{\Gamma}\right)\boldsymbol{\delta}_{\rm E}^{k+1}-\rho\boldsymbol{\Gamma}^{\rm T}\left(\mathbf{c}+\hat{\mathbf{w}}^{k+1}+\frac{\hat{\boldsymbol{\lambda}}^{k}}{\rho}\right)=\mathbf{0}
\end{align}
Substituting \eqref{equation2-30} into \eqref{equation2-47} we have
\begin{align}
\label{equation2-48}
2\mathbf{U}\boldsymbol{\delta}_{\rm E}^{k+1}-\boldsymbol{\Gamma}^{\rm T}\hat{\boldsymbol{\lambda}}^{k+1}=\mathbf{0}
\end{align}
which means
\begin{align}
\label{equation2-49}
&\left\|\hat{\boldsymbol{\lambda}}^{k+1}-\hat{\boldsymbol{\lambda}}^{k}\right\|^{2}_{2}\notag\\
&=\left\|2\mathbf{U}\left(\boldsymbol{\delta}_{\rm E}^{k+1}-\boldsymbol{\delta}_{\rm E}^{k}\right)\right\|^{2}_{2}\notag\\
&\leq 4\varphi^{2}\left\|\boldsymbol{\delta}_{\rm E}^{k+1}-\boldsymbol{\delta}_{\rm E}^{k}\right\|^{2}_{2}
\end{align}
where $\varphi={\rm eig_{max}}\left(\mathbf{U}\right)$.

Substituting \eqref{equation2-49} into \eqref{equation2-46} results in
\begin{align}
\label{equation2-50}
&\hat{L}_{\rho}\left(\boldsymbol{\delta}_{\rm E}^{k+1},\hat{\boldsymbol{\omega}}^{k+1},\hat{\boldsymbol{\lambda}}^{k+1}\right)\notag\\
&\leq \hat{L}_{\rho}\left(\boldsymbol{\delta}_{\rm E}^{k},\hat{\boldsymbol{\omega}}^{k},\hat{\boldsymbol{\lambda}}^{k}\right)-\left(\frac{\rho}{2}-\frac{4\varphi^{2}}{\rho}\right)\left\|\boldsymbol{\delta}_{\rm E}^{k+1}-\boldsymbol{\delta}_{\rm E}^{k}\right\|^{2}_{2}
\end{align}
 which means if the condition
\begin{align}
\label{equation2-51}
\rho> 2\sqrt{2}\varphi
\end{align}
holds, $\hat{L_{\rho}}$ is monotonously decreasing in the iteration procedure.

\subsection{Complexity Analysis for the Second ADMM Scheme}
We analyze the computational complexity of the proposed algorithm in terms of the number of real multiplication operations.

We can take $\rho=1$.
Considering the special structure of $\boldsymbol{\Gamma}$ shown in \eqref{equation2-25} and $\boldsymbol{\Gamma}^{\rm T}\boldsymbol{\Gamma}$ shown in \eqref{equation2-40}, the computational complexity can be greatly reduced. 
In line 4 of Algorithm 2, the computation of $\left(2\mathbf{U}+\boldsymbol{\Gamma}^{\rm T}\boldsymbol{\Gamma}\right)^{-1}$ requires $\frac{1}{3}\left(2NK\right)^{3}$ real multiplications, which involve the cost of computing the Cholesky factorization.
The computation of the matrix product of $\left(2\mathbf{U}+\boldsymbol{\Gamma}^{\rm T}\boldsymbol{\Gamma}\right)^{-1}\boldsymbol{\Gamma}^{\rm T}\left(\mathbf{c}+\hat{\mathbf{w}}^{k}+\hat{\boldsymbol{\lambda}}^{k}\right)$ requires $2NK\left(2NK+1\right)$ real multiplications. The total cost of $\boldsymbol{\delta}_{\rm E}$-update is $\left(\frac{1}{3}\left(2NK\right)^{3}+2NK\left(2NK+1\right)\right)$ real multiplications.
The cost of projection $\hat{\boldsymbol{\omega}}$-update in line 5 is negligible. And the $\hat{\boldsymbol{\lambda}}$-update in line 6 does not require any real multiplication.
Therefore, Algorithm 2 requires $\left(\frac{1}{3}\left(2NK\right)^{3}+2NK\left(2NK+1\right)\right)$ real multiplications when $k = 1$.

Given that $\left(2\mathbf{U}+\boldsymbol{\Gamma}^{\rm T}\boldsymbol{\Gamma}\right)^{-1}$ does not change in each iteration, we can cache the result to perform the subsequent iterations efficiently. Accordingly, Algorithm 2 requires $2NK\left(2NK+1\right)$ real multiplications for each iteration when $k \geq 2$. 
The total number of real multiplications for Algorithm 2 is $\left[\frac{1}{3}\left(2NK\right)^{3}+T\cdot 2NK\left(2NK+1\right)\right]$, where $T$ denotes the number of iterations.

\section{Simulation Result}
we assume standard Rayleigh fading channel, random complex Gaussian distributed noise. The signal transmitting power is $p_{0}=1$, and the SNR is defined as $\frac{1}{\sigma^{2}}$, where $\sigma^{2}$ is the noise power. The execution time results are obtained from a Windows 11 Desktop with i9-10900 and 16GB RAM.

For clarity, the following abbreviations are used throughout this section:

\begin{enumerate}
\item{ZF: Traditional ZF precoding with block-level power normalization;}
\item{RZF: Traditional ZF precoding with block-level power normalization;}
\item{CI-SLP-QP: Traditional CI-SLP method solved by \emph{quadprog} in \emph{matlab}, $\mathcal{P}_{8}$ in \cite{22};}
\item{CI-BLP-CVX: CI-BLP method solved by \emph{CVX} in \emph{matlab}, $\mathcal{P}_{2}$;}
\item{CI-BLP-IPM: CI-BLP method solved by IPM, $\mathcal{P}_{2}$;}
\item{CI-BLP-QP: CI-BLP method solved by \emph{quadprog} in \emph{matlab}, $\mathcal{P}_{2}$;}
\item{CI-BLP-ADMM-P2-($K_{max}$): CI-BLP method solved by the proposed ADMM algorithm based on $\mathcal{P}_{2}$ with the maximum number of iterations $K_{max}$;}
\item{CI-BLP-ADMM-P5-($K_{max}$): CI-BLP method solved by the proposed ADMM algorithm based on $\mathcal{P}_{5}$ with the maximum number of iterations $K_{max}$;}
\end{enumerate}

\begin{figure}[h]
\centering
\includegraphics[width=3.5in]{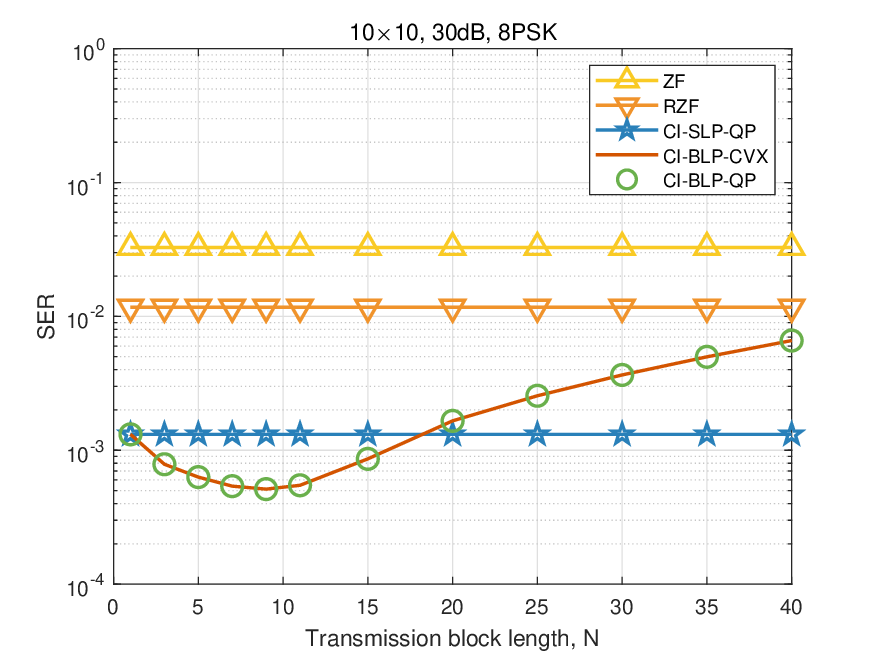}
\caption{SER performance of different schemes, $N_{t}=K=10$, 8PSK, 30dB.}
\label{fig_3}
\end{figure}

To verify the closed-form expression of the CI-BLP optimal precoding matrix for $N< K$, in Fig. 3 we depict the SER with respect to the block length $N$, where 8PSK modulation is employed at a transmit SNR of 30dB, $N_{t}=K=10$. As can be observed, CI-BLP-QP returns the same SER performance as CI-BLP-CVX, which proves that our scenario expansion is feasible.
As the block length $N$ increases, we observe that the SER performance of CI-BLP firstly improves since the benefit of the relaxed power constraint, while the SER performance of CI-BLP becomes worse as $N$ further increases, because the benefit of the relaxed power constraint cannot further compensate for the loss of the fixed precoder.
This also shows that it is meaningful to extend the research of low complexity solution of CI-BLP to $N< K$ scenarios.

\begin{figure}[h]
\centering
\subfloat[]{\includegraphics[width=1.65in]{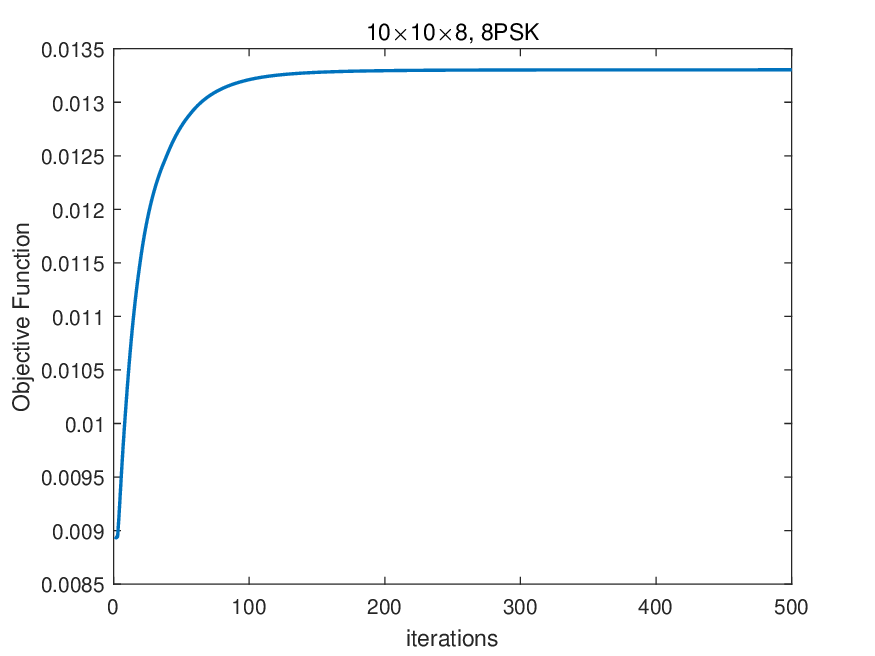}%
\label{fig_first_case}}
\hfil
\subfloat[]{\includegraphics[width=1.65in]{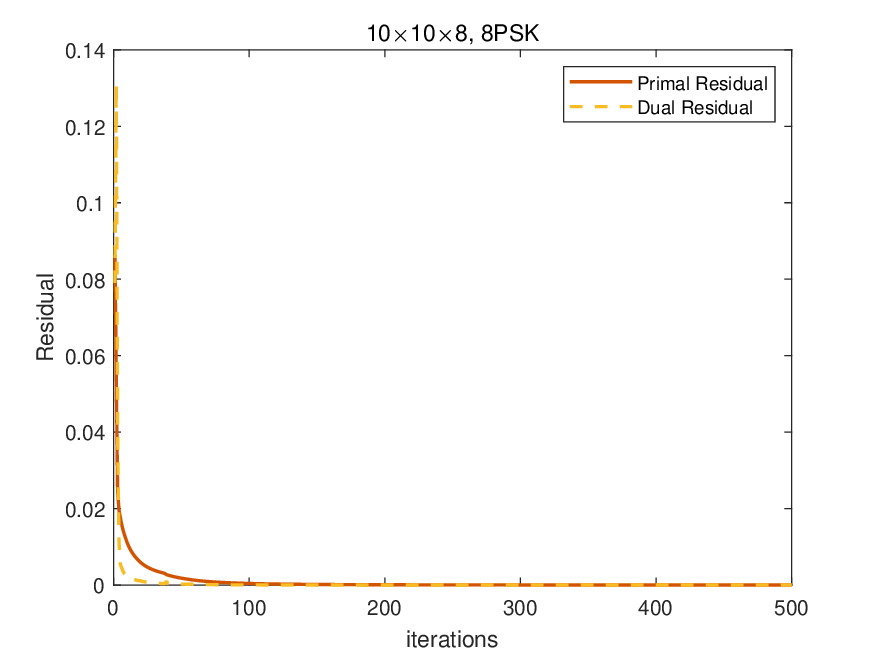}%
\label{fig_second_case}}
\caption{Convergence behavior of proposed ADMM algorithm based on $\mathcal{P}_{5}$, $N_{t}=K=10$, $N=8$, 8PSK. (a) Objective value. (b) Primal residual and dual residual.}
\label{fig_4}
\end{figure}

For general ADMM iterative convergence, it is necessary to analyze the objective value, primal residual and dual residual. For the $k$-th iteration, the objective value refers to $\left[\left(\boldsymbol{\delta}_{\rm E}^{k}\right)^{\rm T}\mathbf{U}\boldsymbol{\delta}_{\rm E}^{k}\right]$. The primal residual refers to $\left[-\boldsymbol{\Gamma}\boldsymbol{\delta}_{\rm E}^{k}+\mathbf{c}+\boldsymbol{\hat{\omega}}^{k}\right]$ and the dual residual refers to $\left[-\boldsymbol{\Gamma}^{\rm T}\mathbf{I}_{\left(2NK+1\right)\times\left(2NK+1\right)}\left(\boldsymbol{\hat{\omega}}^{k}-\boldsymbol{\hat{\omega}}^{k-1}\right)\right]$.
We take the 2-norm as the result of residuals and obtained Fig. 4. In Fig. 4, (a) and (b) show the change of objective value and the change of residual with the number of iterations, respectively.

\begin{figure}[h]
\centering
\includegraphics[width=3.5in]{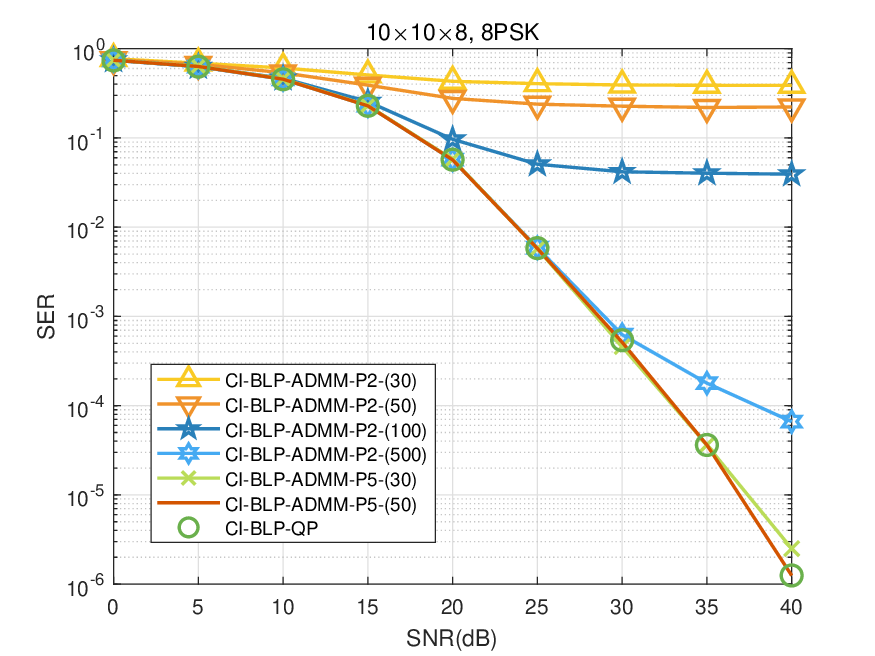}
\caption{SER performance of different schemes, $N_{t}=K=10$, 8PSK, 30dB.}
\label{fig_5}
\end{figure}

Fig. 5 compares the SER performance of the ADMM algorithm proposed based on $\mathcal{P}_{2}$ and the ADMM algorithm proposed based on $\mathcal{P}_{5}$  for a $10\times10$ MU-MISO system with 8PSK modulation, where the length of the block is $N = 8$.
It can be observed that a larger number of iterations corresponds to a more accurate optimal solution.
The ADMM algorithm proposed based on $\mathcal{P}_{2}$ does not achieve satisfactory results although the maximum number of iterations is increased to 500.
However, the ADMM algorithm based on $\mathcal{P}_{5}$ can achieve satisfactory results after 50 or even 30 iterations.
This benefit is the result of our equivalent transformation of $\mathcal{P}_{2}$.

\begin{figure}[h]
\centering
\includegraphics[width=3.5in]{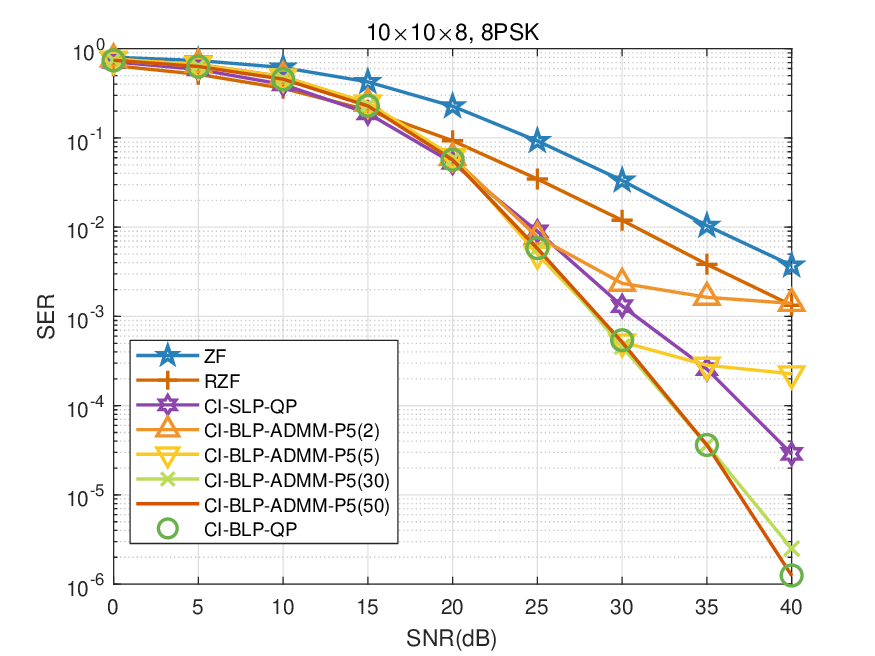}
\caption{SER performance of different schemes, $N_{t}=K=10$, 8PSK, 30dB.}
\label{fig_6}
\end{figure}

Fig. 6 depicts the SER of the proposed CI-BLP scheme when 8PSK modulation is employed in a $10\times10$ MU-MISO system, where the length of the block is $N = 8$. As can
be observed, both CI-based precoding approaches achieve an improved performance over ZF precoding. 
When the length of the block is $N = 8$, we observe that CI-BLP offers noticeable performance gains over traditional CI-SLP, owing to the relaxed power constraint over the entire block.
We observe that a flexible trade-off between SER performance and computational complexity can be achieved by selecting different maximum iterations for the proposed ADMM algorithm.

\begin{figure}[h]
\centering
\includegraphics[width=3.5in]{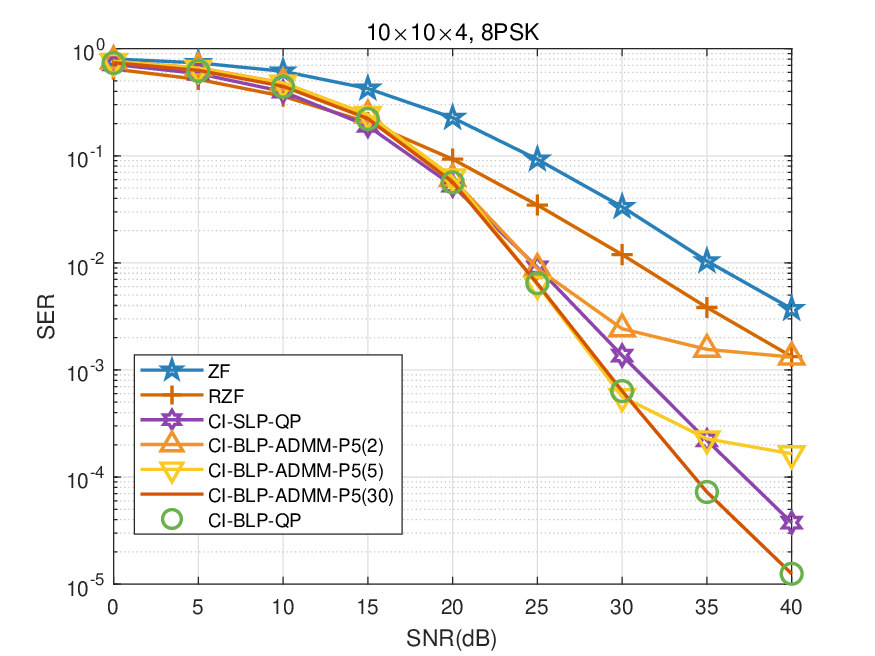}
\caption{SER performance of different schemes, $N_{t}=K=10$, 8PSK, 30dB.}
\label{fig_7}
\end{figure}

Fig. 7 compares the SER performance of different CI precoding approaches for a $10\times10$ MU-MISO system with 8PSK modulation, where the length of the block is decreased to $N = 4$. In this case, while we observe that the performance gain of CI-BLP becomes less significant, it still outperforms traditional CI-SLP.
Due to the smaller problem size, the number of iterations required by the proposed ADMM algorithm is further reduced.

\begin{figure}[h]
\centering
\includegraphics[width=3.5in]{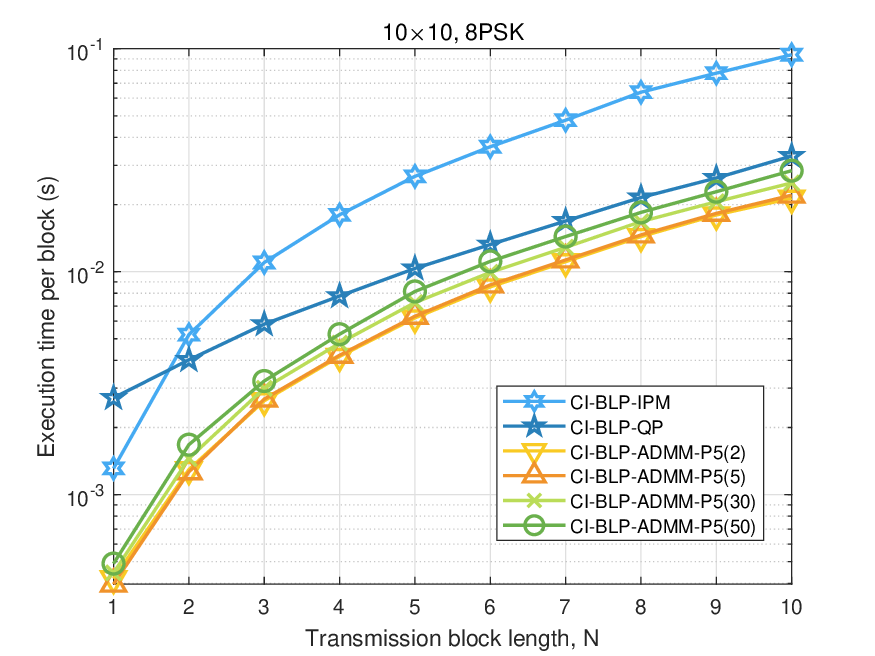}
\caption{SER performance of different schemes, $N_{t}=K=10$, 8PSK, 30dB.}
\label{fig_8}
\end{figure}

in Fig. 8 we compare the execution time required for each scheme as an indication to show the potential complexity benefits of the proposed ADMM algorithm.
It is observed that the proposed ADMM algorithm is much faster than traditional IPM. More importantly, our proposed ADMM algorithm is more time-efficient than \emph{quadprog}, which motivates the use of the block-level CI beamforming in practice.

\section{Conclusion}
In this paper, we focus on extending the analysis on CI-BLP to the case where
the number of symbol slots in a block is smaller than the number of users and obtain a QP optimization on simplex in this case. We study the possibility of applying the iterative closed-form algorithm for QP problem in CI-SLP to CI-BLP. We further propose a new ADMM algorithm. All subproblems have simple closed-form solutions.
The proposed algorithm is shown to achieve an identical performance to \emph{quadprog} with a reduced computational cost, which enables the use of block-level CI precoding in
practical wireless systems.

\begin{appendices}
\section{Proof for Proposition 1}
By the definition of eigenvalues, for any $m \in \left\{r_{g}+1,r_{g}+2,\cdots,M_{g}\right\}$,
\begin{align}
\label{equation44}
\mathbf{G}\cdot\mathbf{v}_{\rm G}^{m}=0\cdot\mathbf{v}_{\rm G}^{m}=\mathbf{0}
\end{align}
By substituting 
\begin{align}
\label{equation45}
\mathbf{G}&=\mathbf{s}_{\rm G}^{1}\cdot\left(\mathbf{s}_{\rm G}^{1}\right)^{\rm T}+\mathbf{s}_{\rm G}^{2}\cdot\left(\mathbf{s}_{\rm G}^{2}\right)^{\rm T}+\cdots+\mathbf{s}_{\rm G}^{N_{g}}\cdot\left(\mathbf{s}_{\rm G}^{N_{g}}\right)^{\rm T}
\end{align}
into \eqref{equation44}, we can obtain
\begin{align}
\label{equation46}
\mathbf{s}_{\rm G}^{1}\cdot\left(\mathbf{s}_{\rm G}^{1}\right)^{\rm T}\cdot\mathbf{v}_{\rm G}^{m}+&\mathbf{s}_{\rm G}^{2}\cdot\left(\mathbf{s}_{\rm G}^{2}\right)^{\rm T}\cdot\mathbf{v}_{\rm G}^{m}+\cdots \notag \\
&+\mathbf{s}_{\rm G}^{N_{g}}\cdot\left(\mathbf{s}_{\rm G}^{N_{g}}\right)^{\rm T}\cdot\mathbf{v}_{\rm G}^{m}=\mathbf{0}
\end{align}
Since $\mathbf{s}_{\rm G}^{1},\mathbf{s}_{\rm G}^{2},\cdots,\mathbf{s}_{\rm G}^{r_{g}}$ is a maximal system of linear independence, we can get the following expression
\begin{align}
\label{equation47}
\mathbf{s}_{\rm G}^{r_{g}+1}=k_{1,1}\mathbf{s}_{\rm G}^{1}+&k_{1,2}\mathbf{s}_{\rm G}^{2}+\cdots+k_{1,r_{g}}\mathbf{s}_{\rm G}^{r_{g}}\notag\\
\mathbf{s}_{\rm G}^{r_{g}+2}=k_{2,1}\mathbf{s}_{\rm G}^{1}+&k_{2,2}\mathbf{s}_{\rm G}^{2}+\cdots+k_{2,r_{g}}\mathbf{s}_{\rm G}^{r_{g}}\notag\\
&\vdots\notag\\
\mathbf{s}_{\rm G}^{N_{g}}=k_{N_{g}-r_{g},1}\mathbf{s}_{\rm G}^{1}+&k_{N_{g}-r_{g},2}\mathbf{s}_{\rm G}^{2}+\cdots+k_{N_{g}-r_{g},r_{g}}\mathbf{s}_{\rm G}^{r_{g}}\notag\\
\end{align}
By substituting \eqref{equation47} into \eqref{equation46}, we can obtain
\begin{align}
\label{equation48}
\mathbf{0}=&\mathbf{s}_{\rm G}^{1}\cdot\left(\mathbf{s}_{\rm G}^{1}\right)^{\rm T}\cdot\mathbf{v}_{\rm G}^{m}+\mathbf{s}_{\rm G}^{2}\cdot\left(\mathbf{s}_{\rm G}^{2}\right)^{\rm T}\cdot\mathbf{v}_{\rm G}^{m}+\cdots \notag \\
&+\mathbf{s}_{\rm G}^{N_{g}}\cdot\left(\mathbf{s}_{\rm G}^{N_{g}}\right)^{\rm T}\cdot\mathbf{v}_{\rm G}^{m}\notag \\
=&\mathbf{s}_{\rm G}^{1}\cdot\left(\mathbf{s}_{\rm G}^{1}\right)^{\rm T}\cdot\mathbf{v}_{\rm G}^{m}+\mathbf{s}_{\rm G}^{2}\cdot\left(\mathbf{s}_{\rm G}^{2}\right)^{\rm T}\cdot\mathbf{v}_{\rm G}^{m}+\cdots\mathbf{s}_{\rm G}^{r_{g}}\cdot\left(\mathbf{s}_{\rm G}^{r_{g}}\right)^{\rm T}\cdot\mathbf{v}_{\rm G}^{m}\notag \\
&+\left(k_{1,1}\mathbf{s}_{\rm G}^{1}+k_{1,2}\mathbf{s}_{\rm G}^{2}+\cdots+k_{1,r_{g}}\mathbf{s}_{\rm G}^{r_{g}}\right)\notag\\
&\ \ \ \cdot\left(k_{1,1}\mathbf{s}_{\rm G}^{1}+k_{1,2}\mathbf{s}_{\rm G}^{2}+\cdots+k_{1,r_{g}}\mathbf{s}_{\rm G}^{r_{g}}\right)^{\rm T}\cdot\mathbf{v}_{\rm G}^{m}+\cdots\notag\\
&+\left(k_{N_{g}-r_{g},1}\mathbf{s}_{\rm G}^{1}+k_{N_{g}-r_{g},2}\mathbf{s}_{\rm G}^{2}+\cdots+k_{N_{g}-r_{g},r_{g}}\mathbf{s}_{\rm G}^{r_{g}}\right)\notag\\
&\ \ \ \cdot\left(k_{N_{g}-r_{g},1}\mathbf{s}_{\rm G}^{1}+k_{N_{g}-r_{g},2}\mathbf{s}_{\rm G}^{2}+\cdots+k_{N_{g}-r_{g},r_{g}}\mathbf{s}_{\rm G}^{r_{g}}\right)^{\rm T}\cdot\mathbf{v}_{\rm G}^{m}\notag\\
=&\underbrace{l_{1,1}\mathbf{s}_{\rm G}^{1}}_{\hat{\mathbf{s}}_{\rm G}^{1}}\cdot\left(\mathbf{s}_{\rm G}^{1}\right)^{\rm T}\cdot\mathbf{v}_{\rm G}^{m}+\underbrace{\left(l_{1,2}\mathbf{s}_{\rm G}^{1}+l_{2,2}\mathbf{s}_{\rm G}^{2}\right)}_{\hat{\mathbf{s}}_{\rm G}^{2}}\cdot\left(\mathbf{s}_{\rm G}^{2}\right)^{\rm T}\cdot\mathbf{v}_{\rm G}^{m}+\cdots\notag\\
&+\underbrace{\left(l_{1,r_{g}}\mathbf{s}_{\rm G}^{1}+l_{2,r_{g}}\mathbf{s}_{\rm G}^{2}+\cdots+l_{r_{g},r_{g}}\mathbf{s}_{\rm G}^{r_{g}}\right)}_{\hat{\mathbf{s}}_{\rm G}^{r_{g}}}\cdot\left(\mathbf{s}_{\rm G}^{r_{g}}\right)^{\rm T}\cdot\mathbf{v}_{\rm G}^{m}
\end{align}
where 
\begin{align}
\label{equation49}
l_{i,j}=
\begin{cases}
1+\left(k_{1,i}\right)^{2}+\left(k_{2,i}\right)^{2}+\cdots+\left(k_{N_{g}-r_{g},i}\right)^{2} & \text{if } i=j,\\
2k_{1,i}k_{1,j}+2k_{2,i}k_{2,j}+\cdots+2k_{N_{g}-r_{g},i}k_{N_{g}-r_{g},j} & \text{if } i\neq j.
\end{cases}
\end{align}
Based on \eqref{equation49}, we can obtain that $l_{i,i}\neq 0$. This
means that $\hat{\mathbf{s}}_{\rm G}^{1},\hat{\mathbf{s}}_{\rm G}^{2},\cdots,\hat{\mathbf{s}}_{\rm G}^{r_{g}}$ are linearly independent. Therefore, 
\begin{align}
\label{equation50}
\left(\mathbf{s}_{\rm G}^{1}\right)^{\rm T}\cdot\mathbf{v}_{G}^{m}=\left(\mathbf{s}_{\rm G}^{2}\right)^{\rm T}\cdot\mathbf{v}_{\rm G}^{m}=\cdots=\left(\mathbf{s}_{G}^{r_{g}}\right)^{\rm T}\cdot\mathbf{v}_{\rm G}^{m}=0
\end{align}
Combined with (47), we can further obtain
\begin{align}
\label{equation51}
\left(\mathbf{s}_{\rm G}^{r_{g}+1}\right)&^{\rm T}\cdot\mathbf{v}_{\rm G}^{m}\notag \\
=&\left(k_{1,1}\mathbf{s}_{\rm G}^{1}+k_{1,2}\mathbf{s}_{\rm G}^{2}+\cdots+k_{1,r_{g}}\mathbf{s}_{\rm G}^{r_{g}}\right)^{\rm T}\cdot\mathbf{v}_{\rm G}^{m}\notag\\
=&k_{1,1}\left(\mathbf{s}_{\rm G}^{1}\right)^{\rm T}\cdot\mathbf{v}_{\rm G}^{m}+k_{1,2}\left(\mathbf{s}_{\rm G}^{2}\right)^{\rm T}\cdot\mathbf{v}_{\rm G}^{m}+\cdots \notag \\
&+k_{1,r_{g}}\left(\mathbf{s}_{\rm G}^{r_{g}}\right)^{\rm T}\cdot\mathbf{v}_{\rm G}^{m}\notag\\
=&0\notag\\
\left(\mathbf{s}_{\rm G}^{r_{g}+2}\right)&^{\rm T}\cdot\mathbf{v}_{\rm G}^{m}\notag \\
=&\left(k_{2,1}\mathbf{s}_{\rm G}^{1}+k_{2,2}\mathbf{s}_{\rm G}^{2}+\cdots+k_{2,r_{g}}\mathbf{s}_{\rm G}^{r_{g}}\right)^{\rm T}\cdot\mathbf{v}_{\rm G}^{m}\notag\\
=&k_{2,1}\left(\mathbf{s}_{\rm G}^{1}\right)^{\rm T}\cdot\mathbf{v}_{\rm G}^{m}+k_{2,2}\left(\mathbf{s}_{\rm G}^{2}\right)^{\rm T}\cdot\mathbf{v}_{\rm G}^{m}+\cdots \notag \\
&+k_{2,r_{g}}\left(\mathbf{s}_{\rm G}^{r_{g}}\right)^{\rm T}\cdot\mathbf{v}_{\rm G}^{m}\notag\\
=&0\notag\\
&\ \ \ \vdots\notag \\
\left(\mathbf{s}_{\rm G}^{N_{g}+1}\right)&^{\rm T}\cdot\mathbf{v}_{\rm G}^{m}\notag \\
=&\left(k_{N_{g}-r_{g},1}\mathbf{s}_{\rm G}^{1}+k_{N_{g}-r_{g},2}\mathbf{s}_{\rm G}^{2}+\cdots+k_{N_{g}-r_{g},r_{g}}\mathbf{s}_{\rm G}^{r_{g}}\right)^{\rm T}\cdot\mathbf{v}_{\rm G}^{m}\notag\\
=&k_{N_{g}-r_{g},1}\left(\mathbf{s}_{\rm G}^{1}\right)^{\rm T}\cdot\mathbf{v}_{\rm G}^{m}+k_{N_{g}-r_{g},2}\left(\mathbf{s}_{\rm G}^{2}\right)^{\rm T}\cdot\mathbf{v}_{\rm G}^{m}+\cdots \notag \\
&+k_{N_{g}-r_{g},r_{g}}\left(\mathbf{s}_{\rm G}^{r_{g}}\right)^{\rm T}\cdot\mathbf{v}_{\rm G}^{m}\notag\\
=&0\notag\\
\end{align}
which completes the proof.

\section{Expressions of $\mathbf{U}$, $\hat{\mathbf{U}}$, $\hat{\mathbf{U}}_{1}$ and $\hat{\mathbf{U}}_{2}$}
The new expression of $\mathbf{U}$ is defined as \eqref{equation55}.
The definitions of $\hat{\mathbf{U}}$, $\hat{\mathbf{U}}_{1}$ and $\hat{\mathbf{U}}_{2}$ are \eqref{equation55}, \eqref{equation56} and \eqref{equation57} respectively.
\raggedbottom
\begin{figure*}
    \begin{align}
    \label{equation54}
\mathbf{U}&= 
\begin{bmatrix}
\mathbf{U}_{1,1} & \mathbf{U}_{1,2} & \cdots & \mathbf{U}_{1,N} \\
\mathbf{U}_{2,1} & \mathbf{U}_{2,2} & \cdots & \mathbf{U}_{2,N} \\
\vdots & \vdots & \ddots & \vdots \\
\mathbf{U}_{N,1} & \mathbf{U}_{N,2} & \cdots & \mathbf{U}_{N,N} 
\end{bmatrix}\notag\\
&=
\begin{scriptsize}
\setlength{\arraycolsep}{1.2pt}
\begin{bmatrix}
p_{1,1}\mathbf{A}^{1}\left(\mathbf{A}^{1}\right)^{\rm T}+f_{1,1}\mathbf{A}^{1}\left(\mathbf{B}^{1}\right)^{\rm T}+g_{1,1}\mathbf{B}^{1}\left(\mathbf{A}^{1}\right)^{\rm T}+q_{1,1}\mathbf{B}^{1}\left(\mathbf{B}^{1}\right)^{\rm T} & \cdots & p_{1,N}\mathbf{A}^{1}\left(\mathbf{A}^{N}\right)^{\rm T}+f_{1,N}\mathbf{A}^{1}\left(\mathbf{B}^{N}\right)^{\rm T}+g_{1,N}\mathbf{B}^{1}\left(\mathbf{A}^{N}\right)^{\rm T}+q_{1,N}\mathbf{B}^{1}\left(\mathbf{B}^{N}\right)^{\rm T} \\
p_{2,1}\mathbf{A}^{2}\left(\mathbf{A}^{1}\right)^{\rm T}+f_{2,1}\mathbf{A}^{2}\left(\mathbf{B}^{1}\right)^{\rm T}+g_{2,1}\mathbf{B}^{2}\left(\mathbf{A}^{1}\right)^{\rm T}+q_{2,1}\mathbf{B}^{2}\left(\mathbf{B}^{1}\right)^{\rm T} & \cdots & p_{2,N}\mathbf{A}^{2}\left(\mathbf{A}^{N}\right)^{\rm T}+f_{2,N}\mathbf{A}^{2}\left(\mathbf{B}^{N}\right)^{\rm T}+g_{2,N}\mathbf{B}^{2}\left(\mathbf{A}^{N}\right)^{\rm T}+q_{2,N}\mathbf{B}^{2}\left(\mathbf{B}^{N}\right)^{\rm T} \\
\vdots & \ddots & \vdots \\
p_{N,1}\mathbf{A}^{N}\left(\mathbf{A}^{1}\right)^{\rm T}+f_{N,1}\mathbf{A}^{N}\left(\mathbf{B}^{1}\right)^{\rm T}+g_{N,1}\mathbf{B}^{N}\left(\mathbf{A}^{1}\right)^{\rm T}+q_{N,1}\mathbf{B}^{N}\left(\mathbf{B}^{1}\right)^{\rm T} & \cdots & p_{N,N}\mathbf{A}^{N}\left(\mathbf{A}^{N}\right)^{\rm T}+f_{N,N}\mathbf{A}^{N}\left(\mathbf{B}^{N}\right)^{\rm T}+g_{N,N}\mathbf{B}^{N}\left(\mathbf{A}^{N}\right)^{\rm T}+q_{N,N}\mathbf{B}^{N}\left(\mathbf{B}^{N}\right)^{\rm T} 
\end{bmatrix}
\end{scriptsize}\notag\\
&=
\begin{bmatrix}
\mathbf{I}_{2K \times 2K} & \mathbf{I}_{2K \times 2K} & & & & & \\
 & & \mathbf{I}_{2K \times 2K} & \mathbf{I}_{2K \times 2K} & & & \\
 & & & & \ddots & & \\
 & & & & & \mathbf{I}_{2K \times 2K} & \mathbf{I}_{2K \times 2K}
\end{bmatrix}
\hat{\mathbf{U}}
\begin{bmatrix}
\mathbf{I}_{2K \times 2K} & & & \\
\mathbf{I}_{2K \times 2K} & & & \\
 &\mathbf{I}_{2K \times 2K} & & \\
 &\mathbf{I}_{2K \times 2K} & & \\
 & &\ddots &\\
 & & &\mathbf{I}_{2K \times 2K}\\
 & & &\mathbf{I}_{2K \times 2K}
\end{bmatrix}
    \end{align}

{\noindent}  \rule[-10pt]{17.5cm}{0.05em}\\

\begin{align}
\label{equation55}
\hat{\mathbf{U}}&= 
\begin{scriptsize}
\begin{bmatrix}
p_{1,1}\mathbf{A}^{1}\left(\mathbf{A}^{1}\right)^{\rm T} & f_{1,1}\mathbf{A}^{1}\left(\mathbf{B}^{1}\right)^{\rm T} & p_{1,2}\mathbf{A}^{1}\left(\mathbf{A}^{2}\right)^{\rm T} & f_{1,2}\mathbf{A}^{1}\left(\mathbf{B}^{2}\right)^{\rm T} & \cdots & p_{1,N}\mathbf{A}^{1}\left(\mathbf{A}^{N}\right)^{\rm T} & f_{1,N}\mathbf{A}^{1}\left(\mathbf{B}^{N}\right)^{\rm T} \\
g_{1,1}\mathbf{B}^{1}\left(\mathbf{A}^{1}\right)^{\rm T} & q_{1,1}\mathbf{B}^{1}\left(\mathbf{B}^{1}\right)^{\rm T}& g_{1,2}\mathbf{B}^{1}\left(\mathbf{A}^{2}\right)^{\rm T} & q_{1,2}\mathbf{B}^{1}\left(\mathbf{B}^{2}\right)^{\rm T}& \cdots & g_{1,N}\mathbf{B}^{1}\left(\mathbf{A}^{N}\right)^{\rm T} & q_{1,N}\mathbf{B}^{1}\left(\mathbf{B}^{N}\right)^{\rm T}\\
p_{2,1}\mathbf{A}^{2}\left(\mathbf{A}^{1}\right)^{\rm T} & f_{2,1}\mathbf{A}^{2}\left(\mathbf{B}^{1}\right)^{\rm T} & p_{2,2}\mathbf{A}^{2}\left(\mathbf{A}^{2}\right)^{\rm T} & f_{2,2}\mathbf{A}^{2}\left(\mathbf{B}^{2}\right)^{\rm T} & \cdots & p_{2,N}\mathbf{A}^{2}\left(\mathbf{A}^{N}\right)^{\rm T} & f_{2,N}\mathbf{A}^{2}\left(\mathbf{B}^{N}\right)^{\rm T} \\
g_{2,1}\mathbf{B}^{2}\left(\mathbf{A}^{1}\right)^{\rm T} & q_{2,1}\mathbf{B}^{2}\left(\mathbf{B}^{1}\right)^{\rm T}& g_{2,2}\mathbf{B}^{2}\left(\mathbf{A}^{2}\right)^{\rm T} & q_{2,2}\mathbf{B}^{2}\left(\mathbf{B}^{2}\right)^{\rm T}& \cdots & g_{2,N}\mathbf{B}^{2}\left(\mathbf{A}^{N}\right)^{\rm T} & q_{2,N}\mathbf{B}^{2}\left(\mathbf{B}^{N}\right)^{\rm T}\\
\vdots & \vdots & \vdots & \vdots & \ddots & \vdots & \vdots \\
p_{N,1}\mathbf{A}^{N}\left(\mathbf{A}^{1}\right)^{\rm T} & f_{N,1}\mathbf{A}^{N}\left(\mathbf{B}^{1}\right)^{\rm T} & p_{N,2}\mathbf{A}^{N}\left(\mathbf{A}^{2}\right)^{\rm T} & f_{N,2}\mathbf{A}^{N}\left(\mathbf{B}^{2}\right)^{\rm T} & \cdots & p_{N,N}\mathbf{A}^{N}\left(\mathbf{A}^{N}\right)^{\rm T} & f_{N,N}\mathbf{A}^{N}\left(\mathbf{B}^{N}\right)^{\rm T} \\
g_{N,1}\mathbf{B}^{N}\left(\mathbf{A}^{1}\right)^{\rm T} & q_{N,1}\mathbf{B}^{N}\left(\mathbf{B}^{1}\right)^{\rm T}& g_{N,2}\mathbf{B}^{N}\left(\mathbf{A}^{2}\right)^{\rm T} & q_{N,2}\mathbf{B}^{N}\left(\mathbf{B}^{2}\right)^{\rm T}& \cdots & g_{N,N}\mathbf{B}^{N}\left(\mathbf{A}^{N}\right)^{\rm T} & q_{N,N}\mathbf{B}^{N}\left(\mathbf{B}^{N}\right)^{\rm T}\\
\end{bmatrix}
\end{scriptsize}
\end{align}
{\noindent}  \rule[-10pt]{17.5cm}{0.05em}\\
\begin{align}
\label{equation56}
\hat{\mathbf{U}}_{1}= 
\begin{bmatrix}
p_{1,1} & f_{1,1} & p_{1,2} & f_{1,2} & \cdots & p_{1,N} & f_{1,N}\\
g_{1,1} & q_{1,1} & g_{1,2} & q_{1,2} & \cdots & g_{1,N} & q_{1,N}\\
p_{2,1} & f_{2,1} & p_{2,2} & f_{2,2} & \cdots & p_{2,N} & f_{2,N}\\
g_{2,1} & q_{2,1} & g_{2,2} & q_{2,2} & \cdots & g_{2,N} & q_{2,N}\\
\vdots & \vdots & \vdots & \vdots & \ddots & \vdots & \vdots \\
p_{N,1} & f_{N,1} & p_{N,2} & f_{N,2} & \cdots & p_{N,N} & f_{N,N}\\
g_{N,1} & q_{N,1} & g_{N,2} & q_{N,2} & \cdots & g_{N,N} & q_{N,N}
\end{bmatrix}
\end{align}
\begin{align}
\label{equation57}
\hat{\mathbf{U}}_{2}= 
\begin{bmatrix}
\mathbf{A}^{1}\left(\mathbf{A}^{1}\right)^{\rm T} & \mathbf{A}^{1}\left(\mathbf{B}^{1}\right)^{\rm T} & \mathbf{A}^{1}\left(\mathbf{A}^{2}\right)^{\rm T} & \mathbf{A}^{1}\left(\mathbf{B}^{2}\right)^{\rm T} & \cdots & \mathbf{A}^{1}\left(\mathbf{A}^{N}\right)^{\rm T} & \mathbf{A}^{1}\left(\mathbf{B}^{N}\right)^{\rm T}\\
\mathbf{B}^{1}\left(\mathbf{A}^{1}\right)^{\rm T} & \mathbf{B}^{1}\left(\mathbf{B}^{1}\right)^{\rm T} & \mathbf{B}^{1}\left(\mathbf{A}^{2}\right)^{\rm T} & \mathbf{B}^{1}\left(\mathbf{B}^{2}\right)^{\rm T} & \cdots & \mathbf{B}^{1}\left(\mathbf{A}^{N}\right)^{\rm T} & \mathbf{B}^{1}\left(\mathbf{B}^{N}\right)^{\rm T}\\
\mathbf{A}^{2}\left(\mathbf{A}^{1}\right)^{\rm T} & \mathbf{A}^{2}\left(\mathbf{B}^{1}\right)^{\rm T} & \mathbf{A}^{2}\left(\mathbf{A}^{2}\right)^{\rm T} & \mathbf{A}^{2}\left(\mathbf{B}^{2}\right)^{\rm T} & \cdots & \mathbf{A}^{2}\left(\mathbf{A}^{N}\right)^{\rm T} & \mathbf{A}^{2}\left(\mathbf{B}^{N}\right)^{\rm T}\\
\mathbf{B}^{2}\left(\mathbf{A}^{1}\right)^{\rm T} & \mathbf{B}^{2}\left(\mathbf{B}^{1}\right)^{\rm T} & \mathbf{B}^{2}\left(\mathbf{A}^{2}\right)^{\rm T} & \mathbf{B}^{2}\left(\mathbf{B}^{2}\right)^{\rm T} & \cdots & \mathbf{B}^{2}\left(\mathbf{A}^{N}\right)^{\rm T} & \mathbf{B}^{2}\left(\mathbf{B}^{N}\right)^{\rm T}\\
\vdots & \vdots & \vdots & \vdots & \ddots & \vdots & \vdots \\
\mathbf{A}^{N}\left(\mathbf{A}^{1}\right)^{\rm T} & \mathbf{A}^{N}\left(\mathbf{B}^{1}\right)^{\rm T} & \mathbf{A}^{N}\left(\mathbf{A}^{2}\right)^{\rm T} & \mathbf{A}^{N}\left(\mathbf{B}^{2}\right)^{\rm T} & \cdots & \mathbf{A}^{N}\left(\mathbf{A}^{N}\right)^{\rm T} & \mathbf{A}^{N}\left(\mathbf{B}^{N}\right)^{\rm T}\\
\mathbf{B}^{N}\left(\mathbf{A}^{1}\right)^{\rm T} & \mathbf{B}^{N}\left(\mathbf{B}^{1}\right)^{\rm T} & \mathbf{B}^{N}\left(\mathbf{A}^{2}\right)^{\rm T} & \mathbf{B}^{N}\left(\mathbf{B}^{2}\right)^{\rm T} & \cdots & \mathbf{B}^{N}\left(\mathbf{A}^{N}\right)^{\rm T} & \mathbf{B}^{N}\left(\mathbf{B}^{N}\right)^{\rm T}
\end{bmatrix}
\end{align}
{\noindent}  \rule[-10pt]{17.5cm}{0.05em}\\
\end{figure*}

\section{Proof for Proposition 2}
In order to make the expression more clear, we introduce $\mathbf{x}^{n}$ into this proof:
\begin{align}
\label{equation58}
\mathbf{x}^{n}=\mathbf{s}_{\rm E}^{n},\ \mathbf{x}^{N+n}=\mathbf{c}_{\rm E}^{n}.
\end{align}
Then, $\mathbf{D}$ can be expressed as
\begin{align}
\label{equation59}
\mathbf{D}&=\left[\sum_{n=1}^{N}\mathbf{s}^{n}_{\rm E}\left(\mathbf{s}^{n}_{\rm E}\right)^{\rm T}+\sum_{n=1}^{N}\mathbf{c}^{n}_{\rm E}\left(\mathbf{c}^{n}_{\rm E}\right)^{\rm T}\right] \notag \\
&=\sum_{n=1}^{2N}\mathbf{x}^{n}\left(\mathbf{x}^{n}\right)^{\rm T} \notag \\
&=
\begin{bmatrix}
x^{1}_{1}\cdot\left(\mathbf{x}^{1}\right)^{\rm T}+x^{2}_{1}\cdot\left(\mathbf{x}^{2}\right)^{\rm T}+\cdots+x^{2N}_{1}\cdot\left(\mathbf{x}^{2N}\right)^{\rm T}\\
x^{1}_{2}\cdot\left(\mathbf{x}^{1}\right)^{\rm T}+x^{2}_{2}\cdot\left(\mathbf{x}^{2}\right)^{\rm T}+\cdots+x^{2N}_{2}\cdot\left(\mathbf{x}^{2N}\right)^{\rm T}\\
\vdots\\
x^{1}_{2K}\cdot\left(\mathbf{x}^{1}\right)^{\rm T}+x^{2}_{2K}\cdot\left(\mathbf{x}^{2}\right)^{\rm T}+\cdots+x^{2N}_{2K}\cdot\left(\mathbf{x}^{2N}\right)^{\rm T}\\
\end{bmatrix}
\end{align}
Since $\mathbf{x}^{1},\mathbf{x}^{2},\cdots,\mathbf{x}^{2K}$is linearly independent, the row rank of $\mathbf{D}$ is $\text{min}\left\{2K,2N\right\}$. In the same way we can get that the column rank of $\mathbf{D}$ is also $\text{min}\left\{2K,2N\right\}$, which completes the proof.

\section{Proof for Proposition 3}
According to \eqref{equation25} and \eqref{equation34}, we introduce $\mathbf{x}^{n}$ and $\mathbf{y}^{n}$ into this proof and define them as
\begin{align}
\label{equation60}
\mathbf{x}^{n}&=\mathbf{V}_{\rm D}^{\rm T}\mathbf{s}^{n}_{\rm E},\notag\\
\mathbf{y}^{n}&=\mathbf{V}_{\rm D}^{\rm T}\mathbf{c}^{n}_{\rm E}
\end{align}
Then, $p_{m,n}$, $f_{m,n}$, $g_{m,n}$, $q_{m,n}$ can be written as
\begin{align}
\label{equation61}
p_{m,n}&=\left(\mathbf{s}_{\rm E}^{n}\right)^{\rm T}\mathbf{D}^{+}\mathbf{s}_{\rm E}^{m}\notag\\
&=\left(\mathbf{x}^{n}\right)^{\rm T}\boldsymbol{\varSigma}_{\rm D}^{+}\mathbf{x}^{m}\notag\\
&=\frac{1}{\sigma_{1}}x^{m}_{1}x^{n}_{1}+\frac{1}{\sigma_{2}}x^{m}_{2}x^{n}_{2}+\cdots+\frac{1}{\sigma_{r_{d}}}x^{m}_{r_{d}}x^{n}_{r_{d}}\notag\\
f_{m,n}&=\left(\mathbf{c}_{\rm E}^{n}\right)^{\rm T}\mathbf{D}^{+}\mathbf{s}_{\rm E}^{m}\notag\\
&=\left(\mathbf{y}^{n}\right)^{\rm T}\boldsymbol{\varSigma}_{\rm D}^{+}\mathbf{x}^{m}\notag\\
&=\frac{1}{\sigma_{1}}x^{m}_{1}y^{n}_{1}+\frac{1}{\sigma_{2}}x^{m}_{2}y^{n}_{2}+\cdots+\frac{1}{\sigma_{r_{d}}}x^{m}_{r_{d}}y^{n}_{r_{d}}\notag\\
g_{m,n}&=\left(\mathbf{s}_{\rm E}^{n}\right)^{\rm T}\mathbf{D}^{+}\mathbf{c}_{\rm E}^{m}\notag\\
&=\left(\mathbf{x}^{n}\right)^{\rm T}\boldsymbol{\varSigma}_{\rm D}^{+}\mathbf{y}^{m}\notag\\
&=\frac{1}{\sigma_{1}}y^{m}_{1}x^{n}_{1}+\frac{1}{\sigma_{2}}y^{m}_{2}x^{n}_{2}+\cdots+\frac{1}{\sigma_{r_{d}}}y^{m}_{r_{d}}x^{n}_{r_{d}}\notag\\
q_{m,n}&=\left(\mathbf{c}_{\rm E}^{n}\right)^{\rm T}\mathbf{D}^{+}\mathbf{c}_{\rm E}^{m}\notag\\
&=\left(\mathbf{y}^{n}\right)^{\rm T}\boldsymbol{\varSigma}_{\rm D}^{+}\mathbf{y}^{m}\notag\\
&=\frac{1}{\sigma_{1}}y^{m}_{1}y^{n}_{1}+\frac{1}{\sigma_{2}}y^{m}_{2}y^{n}_{2}+\cdots+\frac{1}{\sigma_{r_{d}}}y^{m}_{r_{d}}y^{n}_{r_{d}}\notag\\
\end{align}
We further introduce $\mathbf{z}^{i}$ into this proof:
\begin{align}
\label{equation62}
\mathbf{z}^{i}=\frac{1}{\sigma_{i}}\left[x^{1}_{i},y^{1}_{i},x^{2}_{i},y^{2}_{i},\cdots,x^{N}_{i},y^{N}_{i}\right],\ i \in \left\{1,\cdots,r_d\right\}
\end{align}
where $i \in \left\{1,\cdots,r_d\right\}$. It is obvious that $\mathbf{z}^{1},\mathbf{z}^{2},\cdots,\mathbf{z}^{r_d}$ is linearly independent and each row of $\hat{\mathbf{U}}_{1}$ can be expressed as a linear sum of $\mathbf{z}^{1},\mathbf{z}^{2},\cdots,\mathbf{z}^{r_d}$. 
\begin{align}
\label{equation63}
\hat{\mathbf{U}}_{1}=
\begin{bmatrix}
x^{1}_{1}\mathbf{z}_{1}+x^{1}_{2}\mathbf{z}_{2}+\cdots+x^{1}_{r_d}\mathbf{z}_{r_d}\\
y^{1}_{1}\mathbf{z}_{1}+y^{1}_{2}\mathbf{z}_{2}+\cdots+y^{1}_{r_d}\mathbf{z}_{r_d}\\
x^{2}_{1}\mathbf{z}_{1}+x^{2}_{2}\mathbf{z}_{2}+\cdots+x^{2}_{r_d}\mathbf{z}_{r_d}\\
y^{2}_{1}\mathbf{z}_{1}+y^{2}_{2}\mathbf{z}_{2}+\cdots+y^{2}_{r_d}\mathbf{z}_{r_d}\\
\vdots\\
x^{N}_{1}\mathbf{z}_{1}+x^{N}_{2}\mathbf{z}_{2}+\cdots+x^{N}_{r_d}\mathbf{z}_{r_d}\\
y^{N}_{1}\mathbf{z}_{1}+y^{N}_{2}\mathbf{z}_{2}+\cdots+y^{N}_{r_d}\mathbf{z}_{r_d}\\
\end{bmatrix}
\end{align}
Thus the row rank of $\hat{\mathbf{U}}_{1}$ is $r_d$. In the same way we can get that the column rank of $\hat{\mathbf{U}}_{1}$ is also $r_d$, which completes the proof.

\section{Proof for Proposition 4}
$\hat{\mathbf{U}}_{2}$ can be decomposed as
\begin{align}
\label{equation64}
\hat{\mathbf{U}}_{2}=\hat{\mathbf{A}}\hat{\mathbf{B}}
\end{align}
where $\hat{\mathbf{A}} \in \mathbb{R}^{4NK \times 2NK}$ and $\hat{\mathbf{B}} \in \mathbb{R}^{2NK \times 4NK}$ are defined as
\begin{align}
\label{equation65}
\hat{\mathbf{A}}=
\begin{bmatrix}
\mathbf{A}^{1} & \mathbf{A}^{1} & \mathbf{A}^{1} & \mathbf{A}^{1} & \cdots & \mathbf{A}^{1} & \mathbf{A}^{1}\\
\mathbf{B}^{1} & \mathbf{B}^{1} & \mathbf{B}^{1} & \mathbf{B}^{1} & \cdots & \mathbf{B}^{1} & \mathbf{B}^{1}\\
\mathbf{A}^{2} & \mathbf{A}^{2} & \mathbf{A}^{2} & \mathbf{A}^{2} & \cdots & \mathbf{A}^{2} & \mathbf{A}^{2}\\
\mathbf{B}^{2} & \mathbf{B}^{2} & \mathbf{B}^{2} & \mathbf{B}^{2} & \cdots & \mathbf{B}^{2} & \mathbf{B}^{2}\\
\vdots & \vdots & \vdots & \vdots & \ddots & \vdots & \vdots & \\
\mathbf{A}^{N} & \mathbf{A}^{N} & \mathbf{A}^{N} & \mathbf{A}^{N} & \cdots & \mathbf{A}^{N} & \mathbf{A}^{N}\\
\mathbf{B}^{N} & \mathbf{B}^{N} & \mathbf{B}^{N} & \mathbf{B}^{N} & \cdots & \mathbf{B}^{N} & \mathbf{B}^{N}\\
\end{bmatrix}
\end{align}
\begin{align}
\label{equation66}
\hat{\mathbf{B}}=
\setlength{\arraycolsep}{1.2pt}
\begin{bmatrix}
\left(\mathbf{A}^{1}\right)^{\rm T} & & & & & & \\
 & \left(\mathbf{B}^{1}\right)^{\rm T} & & & & & \\
 & & \left(\mathbf{A}^{2}\right)^{\rm T} & & & & \\
 & & &\left(\mathbf{B}^{2}\right)^{\rm T} & & & \\
 & & & & \ddots & & & \\
 & & & & & \left(\mathbf{A}^{N}\right)^{\rm T} & \\
 & & & & & & \left(\mathbf{B}^{N}\right)^{\rm T} \\
\end{bmatrix}
\end{align}
It is obvious that the rank of $\hat{\mathbf{A}}$ is $K$ and we can obtain 
\begin{align}
\label{equation67}
\text{rank}\left(\hat{\mathbf{U}}_{2}\right)\leq\text{rank}\left(\hat{\mathbf{A}}\right)=K
\end{align}
From another perspective, $\hat{\mathbf{A}}$ can be decomposed as
\begin{align}
\label{equation68}
\hat{\mathbf{A}}=\hat{\mathbf{U}}_{2}\tilde{\mathbf{B}}
\end{align}
where $\tilde{\mathbf{B}} \in \mathbb{R}^{4NK \times 2NK}$ is defined as \eqref{equation69}.
\begin{figure*}
\begin{align}
\label{equation69}
\tilde{\mathbf{B}}=
\setlength{\arraycolsep}{1.2pt}
\begin{scriptsize}
\begin{bmatrix}
\mathbf{A}^{1}\left(\left(\mathbf{A}^{1}\right)^{\rm T}\mathbf{A}^{1}\right)^{-1} & & & & & & \\
 & \mathbf{B}^{1}\left(\left(\mathbf{B}^{1}\right)^{\rm T}\mathbf{B}^{1}\right)^{-1} & & & & & \\
 & & \mathbf{A}^{2}\left(\left(\mathbf{A}^{2}\right)^{\rm T}\mathbf{A}^{2}\right)^{-1} & & & & \\
 & & & \mathbf{B}^{2}\left(\left(\mathbf{B}^{2}\right)^{\rm T}\mathbf{B}^{2}\right)^{-1} & & & \\
 & & & & \ddots & & & \\
 & & & & & \mathbf{A}^{N}\left(\left(\mathbf{A}^{N}\right)^{\rm T}\mathbf{A}^{N}\right)^{-1} & \\
 & & & & & & \mathbf{B}^{N}\left(\left(\mathbf{B}^{N}\right)^{\rm T}\mathbf{B}^{N}\right)^{-1} \\
\end{bmatrix}
\end{scriptsize}
\end{align}
{\noindent}  \rule[-10pt]{17.5cm}{0.05em}\\
\end{figure*}
\\
Then we can obtain 
\begin{align}
\label{equation70}
\text{rank}\left(\hat{\mathbf{U}}_{2}\right)\geq\text{rank}\left(\hat{\mathbf{A}}\right)=K
\end{align}
With (67), we can get 
\begin{align}
\label{equation71}
\text{rank}\left(\hat{\mathbf{U}}_{2}\right)=\text{rank}\left(\hat{\mathbf{A}}\right)=K
\end{align}
which completes the proof.

\section{Proof for Proposition 5}
First, we introduce $\mathbf{Y} \in \mathbb{R}^{3 \times 3}$ and $\mathbf{X} \in \mathbb{R}^{9 \times 9}$ into this proof and they are defined as
\begin{align}
\label{equation72}
\mathbf{Y}=
\begin{bmatrix}
\mathbf{y}_{1}\\
\mathbf{y}_{2}\\
\mathbf{y}_{3}\\
\end{bmatrix}=
\begin{bmatrix}
y_{1,1} & y_{1,2} & y_{1,3}\\
y_{2,1} & y_{2,2} & y_{2,3}\\
y_{3,1} & y_{3,2} & y_{3,3}
\end{bmatrix}
\end{align}
\begin{align}
\label{equation73}
\mathbf{X}
=
\begin{bmatrix}
\mathbf{x}_{1}\\
\mathbf{x}_{2}\\
\vdots\\
\mathbf{x}_{9}\\
\end{bmatrix}
=
\begin{bmatrix}
\mathbf{X}^{1,1} & \mathbf{X}^{1,2} & \mathbf{X}^{1,3}\\
\mathbf{X}^{2,1} & \mathbf{X}^{2,2} & \mathbf{X}^{2,3}\\
\mathbf{X}^{3,1} & \mathbf{X}^{3,2} & \mathbf{X}^{3,3}
\end{bmatrix}
\end{align}
where $\mathbf{X}^{m,n} \in \mathbb{R}^{3 \times 3}$ is defined as
\begin{align}
\label{equation74}
\mathbf{X}^{m,n}
=
\begin{bmatrix}
\mathbf{x}^{m,n}_{1}\\
\mathbf{x}^{m,n}_{2}\\
\mathbf{x}^{m,n}_{3}\\
\end{bmatrix}
=
\begin{bmatrix}
x^{m,n}_{1,1} & x^{m,n}_{1,2} & x^{m,n}_{1,3}\\
x^{m,n}_{2,1} & x^{m,n}_{2,2} & x^{m,n}_{2,3}\\
x^{m,n}_{3,1} & x^{m,n}_{3,2} & x^{m,n}_{3,3}
\end{bmatrix}
\end{align}
We suppose that $\mathbf{y}_{1}$ and $\mathbf{y}_{2}$ are linearly independent, and $\text{rank}\left(\mathbf{Y}\right)=r_{1}=2$. We also suppose that $\mathbf{x}^{m,n}_{1}$ and $\mathbf{x}^{m,n}_{2}$ are linearly independent, and $\text{rank}\left(\mathbf{X}\right)=\text{rank}\left(\mathbf{X}^{m,n}\right)=r_{2}=2$, which means $\mathbf{x}_{1}$ and $\mathbf{x}_{2}$ are linearly independent.
Then we can get the following expression
\begin{align}
\label{equation75}
\mathbf{y}_{3}=l^{1}\mathbf{y}_{1}+l^{2}\mathbf{y}_{2}
\end{align}
and
\begin{align}
\label{equation76}
\mathbf{x}_{3}=&k_{3}^{1}\mathbf{x}_{1}+k_{3}^{2}\mathbf{x}_{2} \notag \\
\mathbf{x}_{4}=&k_{4}^{1}\mathbf{x}_{1}+k_{4}^{2}\mathbf{x}_{2} \notag \\
&\vdots \notag \\
\mathbf{x}_{9}=&k_{9}^{1}\mathbf{x}_{1}+k_{9}^{2}\mathbf{x}_{2}
\end{align}

We further introduce $\mathbf{Z} \in \mathbb{R}^{9 \times 9}$ and $\tilde{\mathbf{V}} \in \mathbb{R}^{4 \times 9}$, defined as
\begin{align}
\label{equation77}
\mathbf{Z}
=
\begin{bmatrix}
\mathbf{z}_{1}\\
\mathbf{z}_{2}\\
\vdots\\
\mathbf{z}_{9}\\
\end{bmatrix}
=
\begin{bmatrix}
y_{1,1}\mathbf{X}^{1,1} & y_{1,2}\mathbf{X}^{1,2} & y_{1,3}\mathbf{X}^{1,3}\\
y_{2,1}\mathbf{X}^{2,1} & y_{2,2}\mathbf{X}^{2,2} & y_{2,3}\mathbf{X}^{2,3}\\
y_{3,1}\mathbf{X}^{3,1} & y_{3,2}\mathbf{X}^{3,2} & y_{3,3}\mathbf{X}^{3,3}
\end{bmatrix}
\end{align}
\begin{align}
\label{equation78}
\tilde{\mathbf{V}}
=
\begin{bmatrix}
\tilde{\mathbf{v}}_{1}\\
\tilde{\mathbf{v}}_{2}\\
\tilde{\mathbf{v}}_{3}\\
\tilde{\mathbf{v}}_{4}\\
\end{bmatrix}
=
\begin{bmatrix}
y_{1,1}\mathbf{x}_{1}^{1,1} & y_{1,2}\mathbf{x}_{1}^{1,2} & y_{1,3}\mathbf{x}_{1}^{1,3}\\
y_{1,1}\mathbf{x}_{2}^{1,1} & y_{1,2}\mathbf{x}_{2}^{1,2} & y_{1,3}\mathbf{x}_{2}^{1,3}\\
y_{2,1}\mathbf{x}_{1}^{1,1} & y_{2,2}\mathbf{x}_{1}^{1,2} & y_{2,3}\mathbf{x}_{1}^{1,3}\\
y_{2,1}\mathbf{x}_{2}^{1,1} & y_{2,2}\mathbf{x}_{2}^{1,2} & y_{2,3}\mathbf{x}_{2}^{1,3}\\
\end{bmatrix}
\end{align}
With \eqref{equation75} and \eqref{equation76}, each row of $\mathbf{Z}$ can be expressed as a linear combination of $\tilde{\mathbf{v}}_{1},\tilde{\mathbf{v}}_{2},\tilde{\mathbf{v}}_{3},\tilde{\mathbf{v}}_{4}$.
\begin{align}
\label{equation79}
\mathbf{Z}
=
\begin{bmatrix}
\tilde{\mathbf{v}}_{1}\\
\tilde{\mathbf{v}}_{2}\\
k^{1}_{3}\tilde{\mathbf{v}}_{1}+k^{2}_{3}\tilde{\mathbf{v}}_{2}\\
k^{1}_{4}\tilde{\mathbf{v}}_{3}+k^{2}_{4}\tilde{\mathbf{v}}_{4}\\
k^{1}_{5}\tilde{\mathbf{v}}_{3}+k^{2}_{5}\tilde{\mathbf{v}}_{4}\\
k^{1}_{6}\tilde{\mathbf{v}}_{3}+k^{2}_{6}\tilde{\mathbf{v}}_{4}\\
l^{1}k^{1}_{7}\tilde{\mathbf{v}}_{1}+l^{1}k^{2}_{7}\tilde{\mathbf{v}}_{2}+l^{2}k^{1}_{7}\tilde{\mathbf{v}}_{3}+l^{2}k^{2}_{7}\tilde{\mathbf{v}}_{4}\\
l^{1}k^{1}_{8}\tilde{\mathbf{v}}_{1}+l^{1}k^{2}_{8}\tilde{\mathbf{v}}_{2}+l^{2}k^{1}_{8}\tilde{\mathbf{v}}_{3}+l^{2}k^{2}_{8}\tilde{\mathbf{v}}_{4}\\
l^{1}k^{1}_{9}\tilde{\mathbf{v}}_{1}+l^{1}k^{2}_{9}\tilde{\mathbf{v}}_{2}+l^{2}k^{1}_{9}\tilde{\mathbf{v}}_{3}+l^{2}k^{2}_{9}\tilde{\mathbf{v}}_{4}\\
\end{bmatrix}
\end{align}
Next we prove that $\tilde{\mathbf{v}}_{1},\tilde{\mathbf{v}}_{2},\tilde{\mathbf{v}}_{3},\tilde{\mathbf{v}}_{4}$ are linearly independent.

Assuming that $\tilde{\mathbf{v}}_{1},\tilde{\mathbf{v}}_{2},\tilde{\mathbf{v}}_{3},\tilde{\mathbf{v}}_{4}$ are linearly dependent, then there exist coefficients $t_{1},t_{2},t_{3},t_{4}$ that are not all zero, such that
\begin{align}
\label{equation80}
t_{1}\tilde{\mathbf{v}}_{1}+t_{2}\tilde{\mathbf{v}}_{2}+t_{3}\tilde{\mathbf{v}}_{3}+t_{4}\tilde{\mathbf{v}}_{4}=\mathbf{0}^{\rm T}
\end{align}
that is,
\begin{align}
\label{equation81}
&t_{1}y_{1,1}\mathbf{x}^{1,1}_{1}+t_{2}y_{1,1}\mathbf{x}^{1,1}_{2}+t_{3}y_{2,1}\mathbf{x}^{1,1}_{1}+t_{4}y_{2,1}\mathbf{x}^{1,1}_{2} \notag \\
=&\left(t_{1}y_{1,1}+t_{3}y_{2,1}\right)\mathbf{x}^{1,1}_{1}+\left(t_{2}y_{1,1}+t_{4}y_{2,1}\right)\mathbf{x}^{1,1}_{2} \notag \\
=&\mathbf{0}^{\rm T} \notag \\
&t_{1}y_{1,2}\mathbf{x}^{1,2}_{1}+t_{2}y_{1,2}\mathbf{x}^{1,2}_{2}+t_{3}y_{2,2}\mathbf{x}^{1,2}_{1}+t_{4}y_{2,2}\mathbf{x}^{1,2}_{2} \notag \\
=&\left(t_{1}y_{1,2}+t_{3}y_{2,2}\right)\mathbf{x}^{1,2}_{1}+\left(t_{2}y_{1,2}+t_{4}y_{2,2}\right)\mathbf{x}^{1,2}_{2} \notag \\
=&\mathbf{0}^{\rm T} \notag \\
&t_{1}y_{1,3}\mathbf{x}^{1,3}_{1}+t_{2}y_{1,3}\mathbf{x}^{1,3}_{2}+t_{3}y_{2,3}\mathbf{x}^{1,3}_{1}+t_{4}y_{2,3}\mathbf{x}^{1,3}_{2} \notag \\
=&\left(t_{1}y_{1,3}+t_{3}y_{2,3}\right)\mathbf{x}^{1,3}_{1}+\left(t_{2}y_{1,3}+t_{4}y_{2,3}\right)\mathbf{x}^{1,3}_{2} \notag \\
=&\mathbf{0}^{\rm T} \notag \\
\end{align}
Since $\mathbf{x}^{m,n}_{1}$ and $\mathbf{x}^{m,n}_{2}$ are linearly independent, we can get
\begin{align}
\label{equation82}
t_{1}y_{1,1}+t_{3}y_{2,1}=t_{2}y_{1,1}+t_{4}y_{2,1}=0 \notag \\
t_{1}y_{1,2}+t_{3}y_{2,2}=t_{2}y_{1,2}+t_{4}y_{2,2}=0 \notag \\
t_{1}y_{1,3}+t_{3}y_{2,3}=t_{2}y_{1,3}+t_{4}y_{2,3}=0
\end{align}
Since $t_{1},t_{2},t_{3},t_{4}$ are not all zero, the following equation must be satisfied
\begin{align}
\label{equation83}
\frac{y_{1,1}}{y_{2,1}}=\frac{y_{1,2}}{y_{2,2}}=\frac{y_{1,3}}{y_{2,3}}
\end{align}
which contradicts the fact that $\mathbf{y}_{1}$ and $\mathbf{y}_{2}$ are linearly independent. Therefore, the assumption does not hold. $\tilde{\mathbf{v}}_{1},\tilde{\mathbf{v}}_{2},\tilde{\mathbf{v}}_{3},\tilde{\mathbf{v}}_{4}$ are linearly independent. And $\text{rank}\left(\mathbf{Z}\right)=\text{rank}\left(\tilde{\mathbf{V}}\right)=r_{1}\cdot r_{2}$.

We extend this process by substituting $\hat{\mathbf{U}}_{1}$ for $\mathbf{Y}$ and $\hat{\mathbf{U}}_{2}$ for $\mathbf{X}$. And then we can construct a new corresponding $\tilde{\mathbf{V}}$, and we can obtain
\begin{align}
\label{equation84}
\text{rank}\left(\tilde{\mathbf{V}}\right)&=\text{rank}\left(\mathbf{Z}\right) \notag \\
&=\text{rank}\left(\hat{\mathbf{U}}_{1}\right)\cdot\text{rank}\left(\hat{\mathbf{U}}_{2}\right) \notag \\
&=\text{min}\left\{2NK,2K^{2}\right\}
\end{align}
which completes the proof.

\section{Proof for Proposition 6}
we introduce $\hat{\mathbf{X}}=\left[\mathbf{x}_{1},\mathbf{x}_{2},\mathbf{x}_{3},\mathbf{x}_{4},\mathbf{x}_{5},\mathbf{x}_{6}\right]$ into this proof. We suppose that $\text{rank}\left(\hat{\mathbf{X}}\right)=3$ and $\mathbf{x}_{1},\mathbf{x}_{2},\mathbf{x}_{4}$ are linearly independent, which means
\begin{align}
\label{equation85}
\mathbf{x}_{3}=k_{3}^{1}\mathbf{x}_{1}+k_{3}^{2}\mathbf{x}_{2}+k_{3}^{4}\mathbf{x}_{4} \notag \\
\mathbf{x}_{5}=k_{5}^{1}\mathbf{x}_{1}+k_{5}^{2}\mathbf{x}_{2}+k_{5}^{4}\mathbf{x}_{4} \notag \\
\mathbf{x}_{6}=k_{6}^{1}\mathbf{x}_{1}+k_{6}^{2}\mathbf{x}_{2}+k_{6}^{4}\mathbf{x}_{4} 
\end{align}
We further define
\begin{align}
\label{equation86}
\mathbf{X}&=\hat{\mathbf{X}}\cdot
\begin{bmatrix}
\mathbf{I}_{3 \times 3} &\\
& \mathbf{I}_{3 \times 3}
\end{bmatrix} \notag \\
&=\left[\mathbf{x}_{1}+\mathbf{x}_{4},\mathbf{x}_{2}+\mathbf{x}_{5},\mathbf{x}_{3}+\mathbf{x}_{6}\right]
\end{align}
It is obvious that $\text{rank}\left(\mathbf{X}\right) \leq \text{rank}\left(\hat{\mathbf{X}}\right)$. 

Assuming that $\text{rank}\left(\mathbf{X}\right) < \text{rank}\left(\hat{\mathbf{X}}\right)$. Then $\mathbf{x}_{1}+\mathbf{x}_{4},\mathbf{x}_{2}+\mathbf{x}_{5},\mathbf{x}_{3}+\mathbf{x}_{6}$ are linearly dependent, which means there exist coefficients $t_{1},t_{2},t_{3}$ that are not all zero, such that
\begin{align}
\label{equation87}
\mathbf{0}^{\rm T}=&t_{1}\cdot\left(\mathbf{x}_{1}+\mathbf{x}_{4}\right)+t_{2}\cdot\left(\mathbf{x}_{2}+\mathbf{x}_{5}\right)+t_{3}\cdot\left(\mathbf{x}_{3}+\mathbf{x}_{6}\right)\notag\\
=&\left(t_{1}+t_{2}k_{5}^{1}+t_{3}\left(k_{3}^{1}+k_{6}^{1}\right)\right)\cdot\mathbf{x}_{1}\notag \\
&+\left(t_{2}\left(1+k_{5}^{2}\right)+t_{3}\left(k_{3}^{2}+k_{6}^{2}\right)\right)\cdot\mathbf{x}_{2}\notag \\
&+\left(t_{1}+t_{2}k_{5}^{4}+t_{3}\left(k_{3}^{4}+k_{6}^{4}\right)\right)\cdot\mathbf{x}_{4}\notag \\
\end{align}
Since $\mathbf{x}_{1},\mathbf{x}_{2},\mathbf{x}_{4}$ are linearly independent, we can get
\begin{align}
\label{equation88}
t_{1}+t_{2}k_{5}^{1}+t_{3}\left(k_{3}^{1}+k_{6}^{1}\right)=0 \notag \\
t_{2}\left(1+k_{5}^{2}\right)+t_{3}\left(k_{3}^{2}+k_{6}^{2}\right)=0 \notag \\
t_{1}+t_{2}k_{5}^{4}+t_{3}\left(k_{3}^{4}+k_{6}^{4}\right)=0
\end{align}
Since $t_{1},t_{2},t_{3}$ are not all zero, the following equation must be satisfied
\begin{align}
\label{equation89}
\frac{\left(k_{3}^{2}+k_{6}^{2}\right)k_{5}^{1}}{1+k_{5}^{2}}+k_{3}^{1}+k_{6}^{1}=\frac{\left(k_{3}^{2}+k_{6}^{2}\right)k_{5}^{4}}{1+k_{5}^{2}}+k_{3}^{4}+k_{6}^{4}
\end{align}
In general, this strict equality condition in \eqref{equation89} is almost impossible to satisfy, which means the assumption does not hold. Therefore, $\text{rank}\left(\mathbf{X}\right) = \text{rank}\left(\hat{\mathbf{X}}\right)$.
We extend this process by substituting $\hat{\mathbf{U}}$ for $\hat{\mathbf{X}}$ and $\mathbf{U}$ for $\mathbf{X}$. And then we can obtain
\begin{align}
\label{equation90}
\text{rank}\left(\mathbf{U}\right)&=\text{rank}\left(\hat{\mathbf{U}}\right) \notag \\
&=\text{min}\left\{2NK,2K^{2}\right\}
\end{align}
which completes the proof.

\end{appendices}

\bibliographystyle{IEEEtran}
\bibliography{ref}

\vfill

\end{document}